\documentclass[preprint,12pt,authoryear]{elsarticle}

\usepackage{amssymb}
\usepackage{amsmath}
\usepackage{hyperref}
\hypersetup{colorlinks,allcolors=black}
\usepackage{pbox}
\usepackage{subcaption}

\journal{Medical Image Analysis}

\begin{document}

\begin{frontmatter}

\title{Explaining Uncertainty in Multiple Sclerosis Cortical Lesion Segmentation Beyond Prediction Errors}

\author[unil,chuv,hes,cibm]{Nataliia Molchanova\corref{cor1}}
\cortext[cor1]{Corresponding author.}

\author[cibm,chuv,unil]{Pedro M. Gordaliza}

\author[unibas1,unibas2,unibas3,genova]{Alessandro Cagol}
\author[unibas1,unibas2,unibas3]{Mario Ocampo--Pineda}
\author[unibas1,unibas2,unibas3]{Po--Jui Lu}
\author[unibas1,unibas2,unibas3]{Matthias Weigel}
\author[unibas1,unibas2,unibas3]{Xinjie Chen}

\author[sinai]{Erin S. Beck}

\author[nih]{Haris Tsagkas}
\author[nih]{Daniel Reich}

\author[uclouvain]{Anna Stölting}
\author[uclouvain]{Pietro Maggi}

\author[epfl]{Delphine Ribes}

\author[hes]{Adrien Depeursinge}

\author[unibas1,unibas2,unibas3]{Cristina Granziera}

\author[hes,unige]{Henning Müller}

\author[cibm,chuv,unil]{Meritxell Bach Cuadra}

\affiliation[unil]{
organization={Faculty of Biology and Medicine, University of Lausanne (UNIL)}, 
city={Lausanne},
country={Switzerland}
}
\affiliation[chuv]{
organization={Radiology Department, Lausanne University Hospital (CHUV)}, 
city={Lausanne},
country={Switzerland}
}
\affiliation[hes]{
organization={MedGIFT, Institute of Informatics, School of Management, HES--SO Valais--Wallis University of Applied Sciences and Arts Western Switzerland},
city={Sierre},
country={Switzerland}
}
\affiliation[cibm]{
organization={CIBM Center for Biomedical Imaging},
city={Lausanne},
country={Switzerland}
}
\affiliation[unige]{
organization={Department of Radiology and Medical Informatics, University of Geneva},
city={Geneva},
country={Switzerland}
}
\affiliation[unibas1]{
organization={Translational Imaging in Neurology (ThINK) Basel, Department of Medicine and Biomedical Engineering, University Hospital Basel and University of Basel},
city={Basel},
country={Switzerland}
}
\affiliation[unibas2]{
organization={Multiple Sclerosis Center, Department of Neurology, University Hospital Basel},
city={Basel},
country={Switzerland}
}
\affiliation[unibas3]{
organization={Research Center for Clinical Neuroimmunology and Neuroscience Basel (RC2NB), University Hospital Basel and University of Basel},
city={Basel},
country={Switzerland}
}
\affiliation[genova]{
organization={ Dipartimento di Scienze della Salute, Università degli Studi di Genova},
city={Genova},
country={Italy}
}
\affiliation[sinai]{
organization={Department of Neurology, Icahn School of Medicine at Mount Sinai}, 
city={New York City},
country={USA}
}
\affiliation[nih]{
organization={Translational Neuroradiology Section, National Institute of Neurological Disorders and Stroke, National Institutes of Health},
city={Bethesda},
country={USA}
}
\affiliation[uclouvain]{
organization={Neuroinflammation Imaging Lab (NIL), Université catholique de Louvain},
city={Brussels},
country={Belgium}
}
\affiliation[epfl]{
organization={EPFL+ECAL Lab, École polytechnique fédérale de Lausanne (EPFL)},
city={Lausanne},
country={Switzerland}
}

\begin{abstract}
Trustworthy artificial intelligence (AI) is essential in healthcare, particularly for high--stakes tasks like medical image segmentation. Explainable AI and uncertainty quantification significantly enhance AI reliability by addressing key attributes such as robustness, usability, and explainability. 
Despite extensive technical advances in uncertainty quantification for medical imaging, understanding the clinical informativeness and interpretability of uncertainty remains limited.
This study presents an interpretability framework for analyzing lesion-scale predictive uncertainty in cortical lesion segmentation in multiple sclerosis using deep ensembles.
The analysis shifts the focus from the uncertainty--error relationship towards clinically relevant medical and engineering factors.
Our findings reveal that instance--wise uncertainty is strongly related to lesion size, shape, and cortical involvement. 
Expert rater feedback confirms that similar factors impede annotator confidence.
Evaluations conducted on two datasets (206 patients, almost 2000 lesions) under both in--domain and distribution--shift conditions highlight the utility of the framework in different scenarios.

\end{abstract}

\begin{graphicalabstract}
\includegraphics[width=\linewidth]{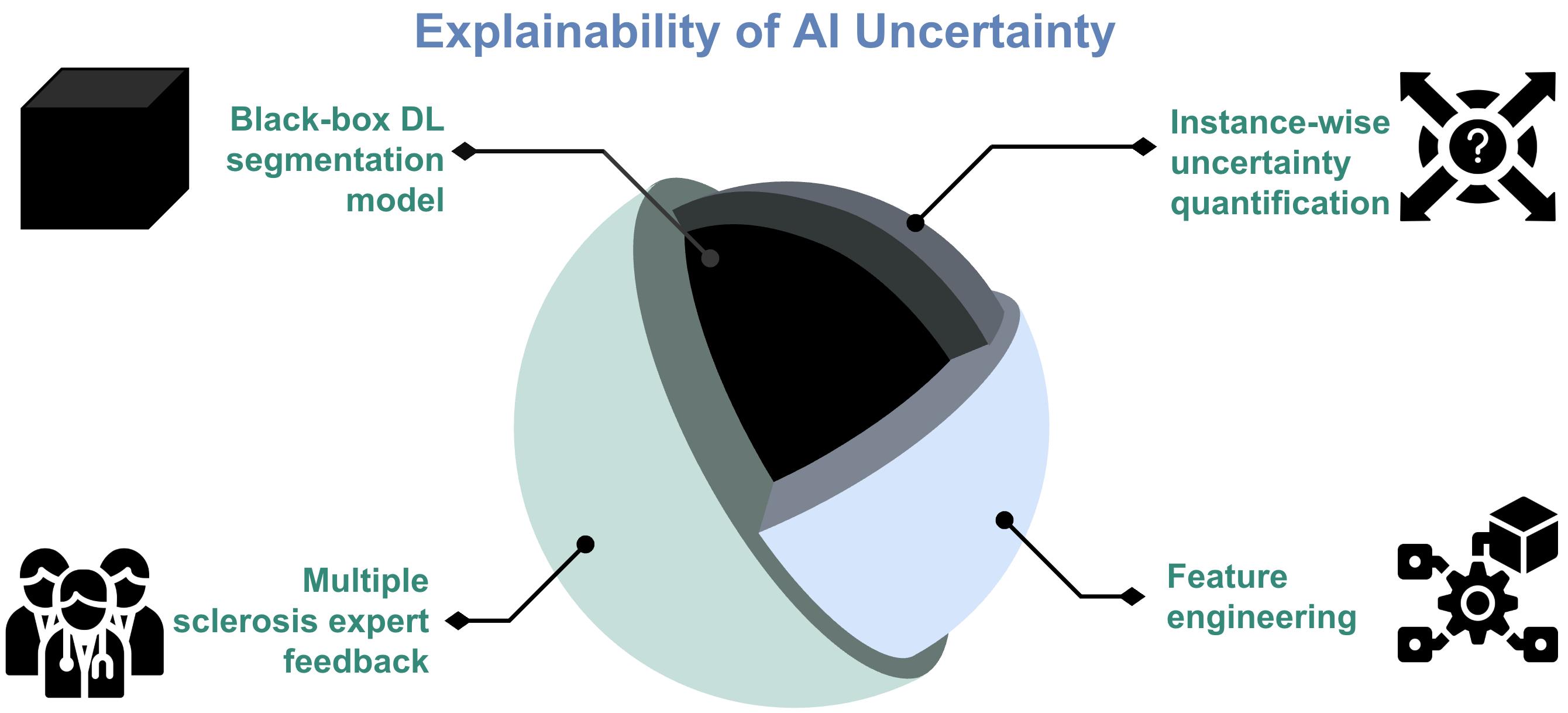}
\end{graphicalabstract}

\begin{highlights}
\item Analysis framework explains lesion-level uncertainty using clinical and imaging features.

\item Uncertainty is linked not only to error but also to meaningful clinical factors.

\item Domain shift affects uncertainty behavior and interpretability.

\item Expert feedback helps assess the clinical relevance of uncertainty explanations.

\item The framework is designed to support interpretation in related uncertainty-aware imaging settings.

\end{highlights}

\begin{keyword}
Uncertainty quantification \sep Instance--wise uncertainty \sep Explained uncertainty \sep Explainable AI \sep Lesion segmentation \sep Multiple sclerosis \sep Magnetic resonance imaging
\end{keyword}

\end{frontmatter}

\section{Introduction}
\label{intro}

Trustworthy artificial intelligence (AI) refers to AI systems designed and implemented to meet essential ethical and performance standards, ensuring safe and beneficial integration into society ~\citep{thiebes_trustworthy_2021, li_trustworthy_2023}. Ensuring AI trustworthiness is crucial in high-risk fields such as healthcare, where AI-driven decisions directly impact patient outcomes ~\citep{albahri_systematic_2023, lambert_trustworthy_2024}. Recent international guidelines for AI in healthcare, FUTURE-AI~\citep{lekadir_future-ai_2025}, have highlighted fairness, universality, traceability, usability, robustness, and explainability as key characteristics needed in the development and deployment of trustworthy AI in healthcare. Regulatory efforts, including the European Union’s AI Act~\citep{noauthor_eu_nodate}, emphasize transparency, accountability, and risk management, mandating clear documentation and interpretability of AI systems to safeguard patient welfare and rights.

Explainable AI (XAI) methods ~\citep{minh_explainable_2022} and uncertainty quantification (UQ) ~\citep{gawlikowski_survey_2023} address several key attributes of trustworthy AI highlighted by international guidelines and regulatory bodies. XAI primarily improves traceability, usability, and explainability by providing insights into how models arrive at specific predictions ~\citep{minh_explainable_2022, lekadir_future-ai_2025}. UQ contributes to robustness, usability, and explainability by quantifying prediction reliability, highlighting potential failure modes, and assisting users in interpreting ambiguous outcomes ~\citep{gawlikowski_survey_2023, lekadir_future-ai_2025}. Together, XAI and UQ facilitate informed human-AI interaction and support human oversight ~\citep{zhang_effect_2020, salvi_explainability_2025}. However, when applied separately, they do not fully address the broader requirements of trustworthy AI~\citep{noauthor_eu_nodate, lekadir_future-ai_2025}.

During the past years, UQ has become increasingly important in medical image analysis, particularly in tasks involving deep learning (DL) based segmentation of pathological features ~\citep{zou_review_2023, lambert_trustworthy_2024, huang_review_2024}. UQ assists clinicians by identifying ambiguous or unreliable predictions and guiding attention to areas that may need manual review. However, effectively evaluating uncertainty is inherently challenging due to the absence of definitive ground truth and the diversity of methods used to quantify uncertainty ~\citep{gawlikowski_survey_2023}. Although substantial technical progress has been achieved, the clinical impact of UQ remains underinvestigated ~\citep{evans_explainability_2022, huet-dastarac_quantifying_2024}. Common evaluation strategies assess the relationship of quantified uncertainty with error but do not reveal whether uncertainty is associated with clinically meaningful factors (e.g., lesion characteristics, image quality) or if it aligns with the expert definition of confidence ~\citep{lambert_trustworthy_2024}. Furthermore, uncertainty is a non-intuitive concept for humans and does not necessarily correlate linearly with annotators' confidence~\citep{huet-dastarac_quantifying_2024}. While the opaque nature of uncertainty hinders clinical adoption, methods that explain the information carried by uncertainty can bridge this gap. Hence, UQ must be evaluated not only for technical performance but also for clinical informativeness and, ideally, for its capacity to support human-AI collaboration~\citep{lee_trust_2004, zhang_effect_2020}.

This work presents an analysis framework layered on top of a standard UQ pipeline to study the clinical informativeness of lesion-scale uncertainty measures in a challenging multiple sclerosis (MS) application~\citep{reich_multiple_2018, thompson_diagnosis_2018}. Automated segmentation of MS lesions is an appropriate setting in which to study the interpretability of uncertainty estimates because recommendations and guidelines already exist for these biomarkers ~\citep{thompson_diagnosis_2018, reich_multiple_2018, hemond_magnetic_2018}. Focal lesion segmentation in MS remains difficult because uncertainty arises from limited training data, distributional shifts, data noise, and ambiguity in pathology definition~\citep{hullermeier_aleatoric_2021, werthen-brabants_role_2025}. This challenge is amplified for recently incorporated biomarkers such as cortical lesions (CLs), paramagnetic rim lesions, and central vein sign, which are small, subtle, and difficult to segment ~\citep{la_rosa_cortical_2022}.

\subsection{Contributions}

This study extends our preliminary MICCAI workshop study \citep{molchanova_interpretability_2025}, which compared lesion-scale uncertainty explanations obtained with Monte Carlo dropout and deep ensembles for CL segmentation. Building on that analysis and prior benchmark evidence that deep ensembles provide the most reliable uncertainty estimates among the methods considered for lesion segmentation, including under distribution shift ~\citep{lambert_trustworthy_2024, malinin_shifts_2022b, molchanova_structural-based_2025}, we focus here on deep ensembles to investigate what the most reliable uncertainty estimates reveal about the clinical interpretability of lesion-scale uncertainty in CL segmentation on MRI in MS. Specifically, the proposed framework:
\begin{enumerate}
    \item Explains the variability in lesion-scale uncertainty in terms of lesion and patient characteristics to unravel the information carried by uncertainty values.

    \item Evaluates uncertainty estimates through their lesion-wise correlation with prediction quality and their behavior in in-domain and distribution-shift settings.

    \item Differentiates between uncertainty patterns that align with clinically meaningful factors and uncertainty that remains less informative for interpretation.
\end{enumerate}

Compared to our earlier study, we introduced the following changes:

\begin{itemize}
    \item \textbf{Clinically meaningful features:} Added lesion-specific and patient-related features to improve the interpretability of uncertainty.

    \item \textbf{Dataset expansion:} The dataset now includes 206 patients and around 2000 lesions from two clinical centers.

    \item \textbf{Domain-shift analysis:} Expanded the evaluation from a single dataset to multiple centers to assess how uncertainty information changes across domains.

    \item \textbf{Transferability of explanations:} Assessed how well uncertainty explanations derived from one dataset apply to another.

    \item \textbf{Ablation study:} Compared regression methods (ordinary least squares, ElasticNet, random forests) and feature combinations for explaining uncertainty variability.

    \item \textbf{Structured expert feedback:} We conducted structured feedback sessions with medical experts to qualitatively assess whether the factors associated with predictive uncertainty align with clinician-reported sources of doubt during CL annotation.
\end{itemize}

\section{Related work}

\subsection{Lesion segmentation in MS}

MS is a chronic inflammatory and neurodegenerative disease with a prevalence of 2 million worldwide, characterized by focal, inflammatory demyelination in the central nervous system ~\citep{reich_multiple_2018, thompson_diagnosis_2018}. MRI plays a key role in MS diagnosis, prognosis, and treatment planning ~\citep{hemond_magnetic_2018}. White matter lesions (WML) detectable on MRI are a hallmark of MS, included in the diagnostic criteria and follow-up pipelines ~\citep{thompson_diagnosis_2018}. DL models have been widely explored to automate the tedious and time-consuming annotation process of WML annotation on MRI ~\citep{commowick_objective_2018, kaur_state---art_2021, spagnolo_how_2023}. Several studies ~\citep{spagnolo_how_2023, rondinella_icpr_2024} have highlighted the utility of U-Net-like architectures ~\citep{ronneberger_u-net_2015, cicek_3d_2016}, dominated by nnU-Net~\citep{isensee_nnu-net_2021, gonzalez_lifelong_2023} in recent years. CLs have emerged as a promising MRI biomarker that supports the differential diagnosis of MS and aids in predicting disease progression ~\citep{thompson_diagnosis_2018, madsen_imaging_2021, la_rosa_cortical_2022, cagol_diagnostic_2024, beck_contribution_2024}. The CL segmentation on MRI presents a greater challenge both for radiologists and machine learning models, given their small size (as small as a few voxels or milliliters), ambiguity in CL definition, and confusion with the WML juxtacortical lesions (see Figure~\ref{fig:examples}). Unlike WML analysis, CL analysis targets a less prevalent but more diagnostically specific lesion type, which changes both the annotation challenge and the clinical interpretation of uncertainty. Due to the lack of publicly available data and the complexity of the task, the automation of the CL segmentation task has not been widely explored. Some articles have proposed joint segmentation of CLs and WMLs ~\citep{fartaria_longitudinal_2019, rosa_multiple_2020}, for which the performance can be dominated by the more prevalent and visible WMLs. CL segmentation was explored in the context of ultra-high field MRI ~\citep{la_rosa_multiple_2022} and compared the visibility of CL on several advanced MRI modalities ~\citep{gordaliza_fluid_2025}.

\subsection{Uncertainty quantification for medical image segmentation}

UQ is widely adopted within the general field of medical image segmentation ~\citep{zou_review_2023, lambert_trustworthy_2024, huang_review_2024}. A range of methodological approaches have been proposed, including Bayesian methods, ensemble-based methods, emerging conformal-prediction approaches, and others (such as test-time augmentations, heteroscedastic models, and uncertainty prediction) ~\citep{zou_review_2023, lambert_trustworthy_2024, huang_review_2024}. Methods such as Monte Carlo dropout~\citep{gal_dropout_2016} and deep ensembles~\citep{lakshminarayanan_simple_2017} dominate the landscape, with ensembles providing superior uncertainty calibration, albeit at a higher computational cost~\citep {lambert_trustworthy_2024}. To mitigate this, single-inference approaches were explored, including temperature-scaled \textit{softmax} probability calibration, evidential DL-based on Dempster–Shafer theory, and learned uncertainty frameworks, predicting uncertainty as one of the outputs ~\citep{lambert_trustworthy_2024}. An important complementary line of work focuses on improving segmentation calibration during training through loss design or regularization. Recent examples include margin-based label smoothing~\citep{murugesan_calibrating_2023}, variational label smoothing~\citep{guo_calseg_2024}, and boundary-weighted logit consistency~\citep{karani_boundary-weighted_2023}. These methods primarily aim to improve the calibration of predictive probabilities and are complementary to the present study, which analyzes the information carried by lesion-scale uncertainty rather than introducing a new calibration strategy.
In parallel, generative models—such as probabilistic U-Nets~\citep{kohl_probabilistic_2018}, PHISeg~\citep{baumgartner_phiseg_2019}, or diffusion-based architectures~\citep{amit_annotator_2023} — were employed to sample diverse plausible segmentation masks, enabling fine-grained modeling of aleatoric uncertainty and inter-rater variability, although often at a higher computational cost. 
Test-time augmentation was proposed within the medical imaging field as a model-agnostic way to quantify aleatoric uncertainty~\citep{wang_aleatoric_2019}.
Several approaches go beyond per-voxel uncertainty and propose to quantify uncertainty per region of interest (ROI), \textit{e.g.}, tumor ~\citep{roy_bayesian_2019} or whole brain ~\citep{jungo_analyzing_2020, molchanova_structural-based_2025}. Commonly, these uncertainty measures average the voxel-wise uncertainty within an ROI~\citep{nair_exploring_2018, roy_bayesian_2019, jungo_analyzing_2020} or quantify the discrepancies in the binary ROI prediction across different samples~\citep{roy_bayesian_2019, molchanova_structural-based_2025}.

For the task of WML MS segmentation, trustworthiness assessment using UQ was previously explored using different techniques, including label-flip uncertainty prediction method ~\citep{mckinley_automatic_2020, lambert_fast_2022}, Monte Carlo dropout ~\citep{nair_exploring_2018, malinin_shifts_2022b}, batch ensembles ~\citep{lambert_fast_2022}, and deep ensembles ~\citep{malinin_shifts_2022b, molchanova_structural-based_2025}. Several studies proposed to quantify uncertainty associated with each predicted lesion region to aid in the false positive lesion discovery~\citep{nair_exploring_2018, molchanova_novel_2023} or segmentation quality prediction~\citep{lambert_beyond_2022}. Further usage of uncertainty varies, including trustworthiness assessment, quality control, data correction, active learning, and out-of-distribution detection ~\citep{zou_review_2023, lambert_trustworthy_2024, huang_review_2024}.

\subsection{Evaluating uncertainty quantification} 

UQ evaluation is an active area of research ~\citep{pignet_legitimate_2024}. Assessing the relationship between uncertainty and error is a common practice, however, it does not unravel all the information brought by the uncertainty values ~\citep{mukhoti_evaluating_2019}. A common evaluation pipeline relies on the assessment of uncertainty calibration (\textit{e.g.}, expected calibration error) ~\citep{guo_calibration_2017} or correlation with error ~\citep{roy_bayesian_2019}, robustness-uncertainty measures (\textit{e.g.}, error retention curves) ~\citep{malinin_shifts_2022a, mehta_qu-brats_2022, molchanova_structural-based_2025}. Some works compare uncertainty against the inter-rater disagreement for segmentation tasks ~\citep{kohl_probabilistic_2018, baumgartner_phiseg_2019}. Qualitative visual assessment of uncertainty maps is also commonly used to inspect whether uncertainty behaves as expected and enhances the interpretability of the uncertainty values. For instance, high uncertainty values are known to appear at the borders of ROIs ~\citep{baumgartner_phiseg_2019, mckinley_automatic_2020, mehta_qu-brats_2022, molchanova_structural-based_2025}. 

Clinical evaluation of uncertainty is limited to the perceptual assessment rather than clinical informativeness analysis ~\citep{evans_explainability_2022, huet-dastarac_quantifying_2024}. A pioneering study ~\citep{evans_explainability_2022} on the XAI and uncertainty perception in the clinics found that trust scores were the most positively perceived explanation type among pathologists, as they provided intuitive and immediately actionable confidence information. A more recent work ~\citep{huet-dastarac_quantifying_2024} studies the visualization scenarios of uncertainty maps in the clinics. This work reported that clinicians preferred binary structure-wise interpretable uncertainty visualizations, paired with voxel-level detail when needed.

Several studies have proposed explainability approaches to unravel the information behind the uncertainty values. 
One study on diabetic retinopathy images classification~\citep{leibig_leveraging_2017} analyzed the causes of uncertainty by examining the predictive variance across Monte Carlo dropout samples and visualizing its behavior to clinical evaluation or unfamiliar inputs, offering qualitative explanations of model uncertainty. Another study~\citep{thiagarajan_explanation_2022} used t-SNE to explain uncertainty in Bayesian CNN-based classification of breast histopathology images, showing how high-uncertainty inputs clustered in latent space and could be flagged for human review, while low-uncertainty cases exhibited significantly improved classification accuracy. In medical image segmentation, our preliminary MICCAI workshop study~\citep{molchanova_interpretability_2025} explored lesion-scale uncertainty explainability for CL segmentation and compared explanations derived from Monte Carlo dropout and deep ensembles. The present manuscript extends that preliminary analysis with a larger two-center setting, a stronger focus on deep ensembles, and structured expert feedback.

\section{Materials and methods}

\subsection{Data}

We used two datasets: i) from a clinical trial acquired at the University Hospital Basel~\citep{granziera_dataset_2018} and ii) a research dataset acquired at the Lausanne University Hospital~\citep{bonnier_dataset_2014} in Switzerland. In total, we included data from 206 patients diagnosed with MS according to McDonald criteria~\citep{thompson_diagnosis_2018}, having different clinical phenotypes: relapsing-remitting (140), primary-progressive (22), and secondary-progressive (44). The dataset comprises 3D magnetization-prepared 2 rapid gradient-echo (MP2RAGE) images~\citep{kober_mp2rage_2012} and manual CL annotations. In the University Hospital Basel, the imaging was performed using a standardized acquisition protocol on a 3 Tesla whole-body MR system (MAGNETOM Prisma, Siemens Healthcare) with a 64-channel head and neck coil. In Lausanne University Hospital, a 3 Tesla whole-body MR system was used (MAGNETOM Trio, Siemens Healthcare) with a 32-channel head and neck coil.

In both medical centers, annotations were obtained by consensus: in University Hospital Basel, by a neurologist and a medical doctor with 5 and 5 years of experience in MRI; in Lausanne University Hospital, by a neurologist and a radiologist with 11 and 7 years of experience in MRI. The identified CLs encompassed both intracortical lesions (confined within the gray matter (GM)) and leukocortical lesions (involving both the cortical GM and the adjacent white matter (WM)). 

\begin{figure}
    \centering
    \caption{Examples of cortical and white matter lesions and their subtypes on MP2RAGE MRI scan: intracortical, leukocortical, juxtacortical, deep white matter lesions. Lesions appear as hypointense regions within GM and WM. The examples illustrate a possible similarity between leukocortical and juxtacortical, contributing to the confusion between cortical and white matter lesions.}
    \includegraphics[width=.95\linewidth]{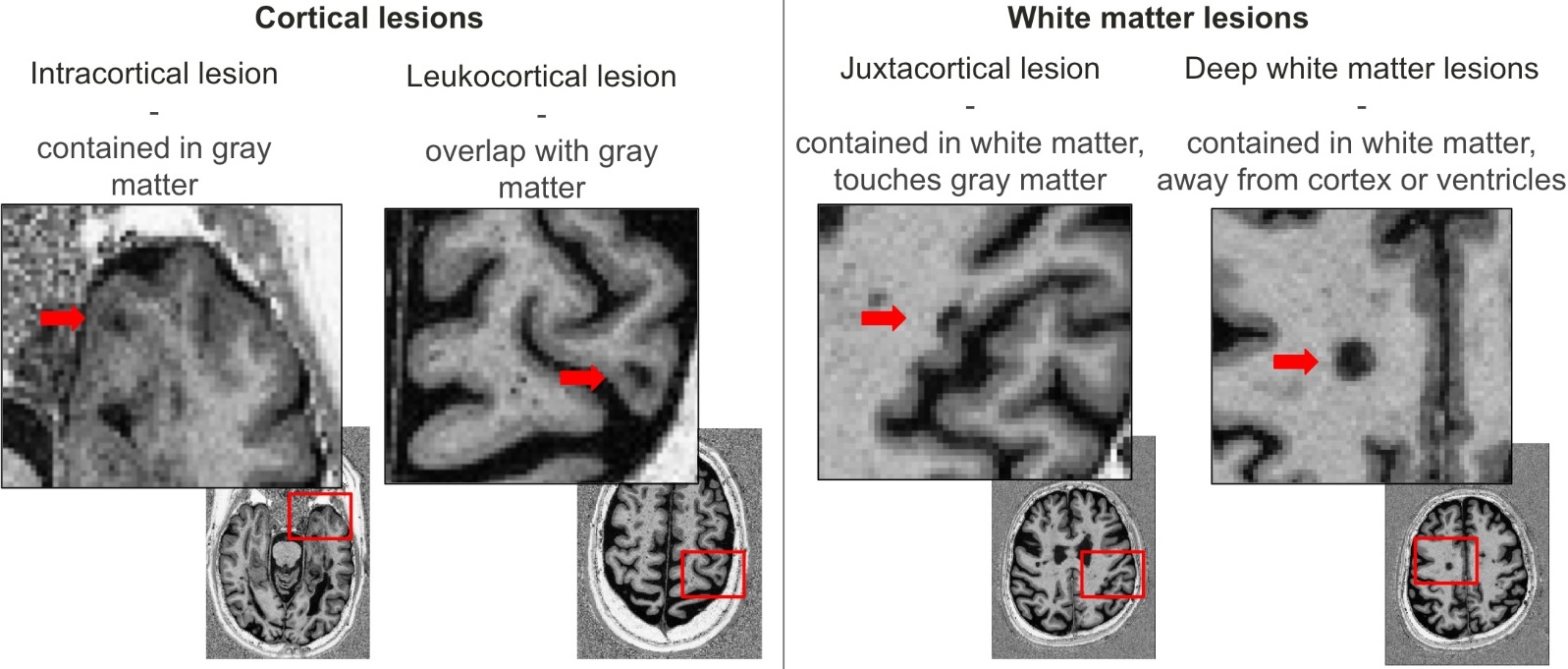}
    \label{fig:cl-wml-ex}
\end{figure}

The data from the University Hospital (n=163) was split into training, validation, and test sets with 109:13:41 subjects respectively, corresponding to 857:69:301 CLs. All train/validation/test splits were performed at the patient (subject) level, and lesions from the same patient never appeared in more than one split. The data from Lausanne University Hospital was solely used for the out-of-distribution evaluation and contains 758 CLs. Further in the manuscript, the in-domain and out-of-domain test sets will be referred to as Test-in and Test-out. The distributional shift between Test-in and Test-out is due to the change of the medical center, MRI scanner model, acquisition protocol, annotators, annotation protocols, and lesion intensity distributions. Examples of CLs are shown in Figure \ref{fig:cl-wml-ex}. The summary of the data information is given in Table \ref{tab:data}.

\begin{table}[h!]
\centering
\caption{Data description. RR - relapsing-remitting, PP - primary-progressive, SP - secondary-progressive; Q2 - median, IQR - interquartile range.}
\small
\begin{tabular}{|p{3cm}|p{5cm}|p{5cm}|}
    \hline
    \textbf{Dataset parameters} & \textbf{In-domain} & \textbf{Out-of-domain} \\
    \hline\hline
    Source & University Hospital Basel & Lausanne University Hospital \\
    \hline
    Disease phenotypes & RR (97), PP (22), SP (44) & RR \\
    \hline
    Number of MS patients & 163 & 43 \\
    \hline
    Total number of CLs & 1227 & 758 \\
    \hline
    M:F ratio & 0.68 & 0.72 \\
    \hline
    Scanner & Magnetom Prisma, Siemens Healthineers & Magnetom Trio, Siemens Healthineers \\
    \hline
    TR, TE, TI1, TI2, ms & 5000, 2.98, 700, 2500 & 5000, 2.89, 700, 2500 \\
    \hline
    Resolution, mm$^3$ & 1.0×1.0×1.0 & 1.0×1.0×1.2 \\
    \hline
    Average lesion intensity Q2 (IQR) & 1006 (879–1130) & 1517 (1256–1758) \\
    \hline
\end{tabular}
\label{tab:data}
\end{table}

\subsection{Model}

We chose nnU-Net, a widely adopted framework for medical image segmentation ~\citep{isensee_nnu-net_2021, gonzalez_lifelong_2023}, previously explored for lesion segmentation specifically ~\citep{rondinella_icpr_2024}. Based on the best practices for medical image segmentation, nnU-Net automatically configures pre-processing, training hyperparameters, and post-processing for a specific dataset. This approach ensures state-of-the-art results without extensive manual intervention, yet it allows for modifications on demand. It is particularly well suited to challenging lesion-segmentation settings with limited data and strong class imbalance, while remaining widely used in recent MS lesion benchmarks ~\citep{rondinella_icpr_2024}. We used the nnU-Net Vanilla architecture operating on 3D full-resolution MP2RAGE scans. For the experiments reported here, all ensemble members used this same architecture and optimization setup; the differences between members arose from independent training runs with different random seeds and the resulting stochastic differences in initialization, data order, and augmentations. Architecture details are given in our GitHub repository.

Given small sizes of CLs and, thus, high class imbalance, the convergence of nnU-Net benefited from adjusting the optimization procedure. The AdamW optimizer was used for all experiments reported in this manuscript, with a learning rate of 0.0001. The learning rate was reduced because the default nnU-Net learning rate did not lead to convergence in our setting, most likely due to the pronounced class imbalance of CL segmentation. AdamW was chosen for its ability to handle sparse gradients and class imbalances, with decoupled weight decay providing improved optimization stability for challenging datasets. We provide the implementations in our GitHub: \url{https://github.com/NataliiaMolch/interpret-lesion-unc}.

\subsubsection{Performance evaluation}

Several measures chosen based on previous works were used to evaluate the model. These measures compare the predicted and ground-truth segmentation: (normalized) Dice similarity score (DSC and nDSC), lesion detection F1-score (LF1), precision (LPPV), and recall (LTPR). All reported evaluations use the patient-level train/validation/test partition defined above. We focused on overlap- and lesion-detection metrics because the study analyzes lesion-wise uncertainty behavior in a setting with many small lesions, for which these measures are more directly aligned with the task than distance-based metrics. Analogously to DSC, nDSC measures the overall segmentation quality, however, it corrects for the difference in lesion loads across subjects, making their comparison fair ~\citep{raina_novel_2023}. For the lesion detection measures, the lesion regions were defined using the connected component analysis with 26 connectivity. True positive lesions were defined as lesions with a non-zero overlap with the ground truth. False positive lesions are predicted connected components without an overlap with the ground truth. False negative lesions are ground truth lesions without an overlap with the predicted lesion mask. The detailed metrics definition is described in \ref{app1}. Mann-Whitney U-tests with Benjamini-Hochberg false discovery rate correction ($\alpha=0.05$) were conducted to check for significant performance changes across test sets.

\subsection{Uncertainty quantification}

\subsubsection{Deep ensemble}

Deep ensembles for UQ have a wide adoption for medical imaging tasks given a strong uncertainty-robustness link and higher calibration compared to other methods, like temperature scaling, test time augmentation, Monte Carlo dropout, or stochastic weight averaging ~\citep{gawlikowski_survey_2023, lambert_trustworthy_2024}. In our setting, deep ensembles were chosen because prior benchmark evidence and our previous lesion-segmentation studies indicated that they provide the most reliable uncertainty estimates among the methods considered, including under distribution shift ~\citep{malinin_shifts_2022b, molchanova_interpretability_2025, molchanova_structural-based_2025}. Deep ensembles are a non-Bayesian UQ method based on training several models with different random seeds. Thus, each model in the ensemble has a distinct weight initialization, augmentations, training examples, shuffling, etc., providing convergence to different local minima. The resulting member weights are treated as sample estimations from the posterior distribution of model parameters. Taken together, the ensemble predictions define a predictive distribution for each voxel rather than a single deterministic segmentation. For each image voxel, uncertainty is computed as a ``spread'' of the member predictions given the same example, \textit{i.e.}, using information theory measures (\textit{e.g.}, variance, mutual information, etc). We implemented UQ using a deep ensemble of $K=5$ models based on the nnU-Net framework architecture, training each model independently without using the bagging technique or any additional diversity-promoting mechanism. This choice provided a practical balance between uncertainty quality and computational cost in our lesion-segmentation experiments.

\subsubsection{Lesion-scale uncertainty measures}

Semantic segmentation task models output a tensor, associating class probabilities with each pixel or voxel of the input image. Such spatial nature allows uncertainty to be quantified at multiple levels: per voxel or ROI (\textit{e.g.}, focal lesions). While pixel-wise uncertainty provides detailed spatial information, instance-level uncertainty (computing a single uncertainty score for each predicted lesion) offers more clinically actionable insights. Computing a single uncertainty value per predicted lesion condenses predictive variability into compact, clinically meaningful information that aligns with how radiologists interpret and report findings, focusing on individual lesions rather than isolated voxels ~\citep{huet-dastarac_quantifying_2024}. 

We quantify the uncertainty associated with each predicted CL using the lesion structural uncertainty (LSU) measure, which we previously proposed for WML segmentation ~\citep{raina_novel_2023}. LSU showed a slightly better ability to capture lesion false discovery errors. Additionally, having a simple, interpretable definition and being bounded from 0 to 1, LSU allows for comparison across different patients. LSU quantifies the average disagreement between each ensemble member (k) lesion prediction, $L^k, k\in \{0, 2, \dots, K-1 \}$, and the ensemble's aggregated prediction $L$. Here, $L$ represents the final lesion segmentation obtained by averaging the probability maps from all $K$ ensemble members and applying a fixed threshold to obtain a binary segmentation mask. Thus, the target of the downstream analysis is lesion-scale uncertainty derived from disagreement within the predictive distribution, not the lesion mask itself. LSU is defined as follows:

\begin{equation*}
    LSU=1 - \frac{1}{K} \sum\limits_{k=0}^{K-1} IoU(L, L^k),
    \label{eq:lsu}
\end{equation*}

where $IoU(\cdot,\cdot)$ denotes the intersection over the union.

We employed a standard evaluation strategy for structure-wise uncertainty measures by computing Spearman’s rank correlation between predicted uncertainty and lesion-wise prediction quality~\citep{roy_bayesian_2019, lambert_beyond_2022}. This approach assesses the extent to which uncertainty correlates with prediction error, based on the premise that higher values from a reliable uncertainty measure should correspond to a greater likelihood of incorrect predictions. Details on the lesion prediction quality metric are provided in the following section.

\subsection{Uncertainty explainability analysis framework}

\begin{figure}[h!]
    \centering
    \caption{Illustration of the proposed framework for explaining instance-wise uncertainty of a DL segmentation model.}
    \includegraphics[width=0.9\linewidth]{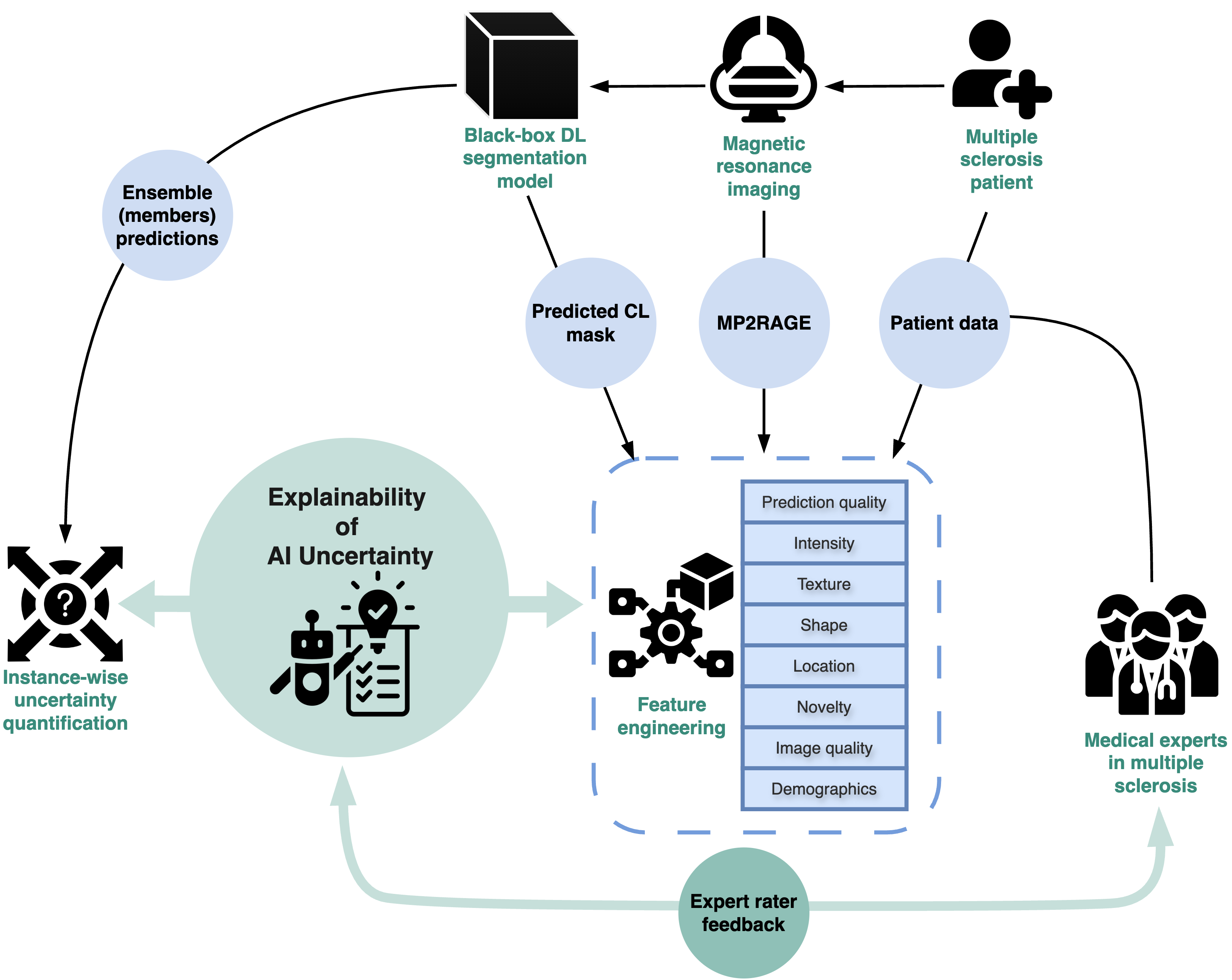}
    \label{fig:diagram}
\end{figure}

We aimed to explain lesion-scale uncertainty values, described by LSU, through their relationship with lesion and patient characteristics. We selected lesion features based on established medical image region characterization practices and clinically relevant metrics, and used them in linear and random forest models to explain uncertainty variability. Feature-importance analysis was then used to examine the information carried by predictive uncertainty across training, in-domain, and out-of-domain settings. We further examined the relationship between important features and the sources of doubt reported by expert raters. Figure \ref{fig:diagram} illustrates the framework. The code is provided in the aforementioned GitHub.

Importantly, the interpretable analysis is not designed to explain why a lesion was segmented as present or absent by a single deterministic model. Instead, it explains why the lesion-scale uncertainty value varies across predicted lesions. In our setting, the response variable is LSU, which is derived from disagreement between the ensemble members and the aggregated segmentation; this differs conceptually from explaining a single nnU-Net segmentation mask, where the target would be lesion presence rather than predictive uncertainty.

\subsubsection{Feature engineering}
\label{sec:feateng}

Each predicted CL was characterized by its segmentation quality, location, intensity, texture, shape, surrounding intensity, and degree of novelty. 

\textbf{Segmentation quality} was computed as an adjusted intersection over the union (IoU\textsubscript{adj})~\citep{rottmann_prediction_2020} between the predicted lesion region ($L$) and the ground truth lesion mask. IoU\textsubscript{adj} is a suitable measure for medical image segmentation, where multiple ground truth instances can overlap with multiple predicted instances. IoU\textsubscript{adj} corrects the union computation, removing the ground truth part explained by other predicted instances. The IoU\textsubscript{adj} computation is detailed in \ref{app2}.

The \textbf{lesion location} was quantified using the MNI atlas~\citep{grabner_symmetric_2006}, separated into the right and left hemispheres. The location features were computed as the intersection between the lesion region and the MNI brain region, divided by the lesion size in voxels. The \textbf{location with respect to the cortex} (\textit{i.e.}, cortex involvement) was computed as the intersection normalized by lesion size with the GM defined by \textit{SynthSeg} ~\citep{billot_synthseg_2023}. In radiological practice, the unclear cortical involvement contributes to the aleatoric uncertainty in distinguishing between CL and WML that only touch the cortex (\textit{i.e.}, juxtocortical lesions).

Radiomics defined by the \textit{PyRadiomics} framework computes \textbf{intensity and shape} features for lesions and their surroundings ~\citep{van_griethuysen_computational_2017}. Perilesional region (\textit{i.e.}, lesion surrounding) was defined as a 4-voxel-wide region around the lesion, which was chosen in agreement with the medical doctors with an admitted degree of ambiguity. We used a morphological dilation with a $3 \times 3 \times 3$ voxels to obtain the lesion surrounding. \textit{PyRadiomics} \textbf{intensity} features (first-order statistics) describe the distribution of voxel intensities within the ROI, such as mean, variance, skewness, and energy, providing insights into the overall brightness, variability, and histogram shape. The \textbf{texture} features (second-order statistics) analyze the spatial relationship and patterns of voxel intensities at specific distances and directions. We focused on the texture features derived from the gray-level co-occurrence matrix (GLCM) with a distance of 1 voxel in all 13 3D directions, quantifying contrast, correlation, energy, and homogeneity. The rest of the texture features are omitted, given the low interpretability and limited number of features below the number of predicted lesions. The \textbf{shape} features capture the 3D geometry of the ROI and focus on the structural and morphological aspects, \textit{e.g.}, volume, surface area, fraction (compactness, sphericity), and longest or shortest axes, among others. 

The \textbf{\textit{degree of novelty}} describes the difference between a lesion and lesions in the Train set. To measure the lesion novelty, we computed the distance between the lesion and the training set using the engineered lesion and perilesional features. For dimensionality and noise reduction, probabilistic principal component analysis (PPCA) ~\citep{bishop_pattern_2006} was applied to the features before computing the distance. PPCA provides a maximum likelihood estimation approach to dimensionality reduction while accounting for noise in the data. We used two measures to compute the distance: Mahalanobis distance and non-zero distance to the closest example in the training set.

\textbf{Patient-wise measures} are also used to examine how global features are related to the lesion uncertainty. First, we added \textit{PyRadiomics} \textbf{intensity} statistics for the whole predicted lesion region. Second, to study the relationship between uncertainty and \textbf{image quality}, we relied on MRIQC~\citep{esteban_mriqc_2017}, characterizing various aspects of MR image quality such as signal-to-noise ratio, motion artifacts, ghosting, and entropy in the whole brain and the specific brain tissues. We selected 13 metrics applicable for structural skull-stripped MRI scans (the selection procedure is detailed in the Supplementary materials). Finally, we used \textbf{patient demographic and clinical information}, such as sex, age, and expanded disability status scale (EDSS). 

The total number of selected features was N=156. Table \ref{tab:features} summarizes all the features.

\begin{table}[h!]
\centering
\caption{Computed features studied in relationship with the lesion-scale uncertainty.}
\renewcommand{\arraystretch}{1.3}
\tiny
    \begin{tabular}{|p{2.2cm}|p{1.3cm}|p{1.5cm}|p{3.5cm}|p{3.5cm}|}
    \hline
    \textbf{Type} & \textbf{Number of features} & \textbf{ROI} & \textbf{Description} & \textbf{Examples} \\
    \hline\hline
    Prediction quality & 1 & lesion & Adjusted intersection over the union between the predicted lesion and the ground truth lesion mask & IoU\textsubscript{adj} \\
    \hline
    Location & 19 & lesion & Intersection between the lesion and atlas-defined brain regions divided by the lesion size & GM overlap, temporal/frontal/occipital left or right lobes overlap \\
    \hline
    Intensity & 18 × 3 & lesion, perilesional region, predicted CL mask
     & Characterize the intensity distribution within ROI & Mean, standard deviation, skewness, percentiles, energy \\
    \hline
    Texture & 24 × 2 & lesion, perilesional region
    & Characterize the distribution of co-occurring voxel values within ROI & Contrast, cluster prominence, entropy \\
    \hline
    Shape & 14 & lesion & Structural and morphological aspects of ROI’s 3D geometry & Volume, surface area, volume-to-surface ratio, flatness, sphericity, maximum diameter \\
    \hline
    Novelty & 2 & lesion & Distance to the training set lesions & Mahalanobis distance or distance to the closest training example \\
    \hline
    Image quality & 13 & brain, GM, WM & MRI quality control measures & Signal-to-noise ratio in brain and brain tissues, intensity non-uniformity (INU) median, entropy focus criterion (EFC) \\
    \hline
    Demographics & 5 & patient & Clinically relevant features & Age, male sex, female sex, MS duration, EDSS score\\
    \hline
    \end{tabular}
    \label{tab:features}
\end{table}

\subsubsection{Feature processing}

The computed feature table contained missing values in the Test-in and Test-out sets. The missing values appeared due to the absence of demographic information (EDSS score missing for 1 patient in Test-in and the second timepoint within Test-out; disease duration missing for 3 patients in Test-out) or due to \textit{PyRadiomics} not producing the results (1 lesion in Test-in and 2 lesions in Test-out). Missing values were imputed using a k-nearest-neighbors approach with five neighbors and uniform weights. The imputation was done before fitting the uncertainty regression model. While imputation before model fitting would typically risk data leakage in predictive tasks, such an approach is beneficial for explainability purposes, providing more reliable estimates of missing values. Before fitting an explainer model, features with low variance were filtered out. The $10^{-6}$ variance threshold was used based on the 32-bit float number precision. The features were standardized separately for each dataset, which makes the coefficients of the linear model comparable to each other.

\subsubsection{Uncertainty regression model}

We explored both linear regression and random forest models to study the relationship between uncertainty (dependent variable) and relevant lesion/patient features (independent variables). On one hand, linear regression provides a simple and interpretable framework with clear insights into the relationship between variables. On the other hand, random forests can model complex and non-linear relationships between independent and dependent variables, albeit at the expense of explainability. 

\textbf{Linear model.} We chose \textit{Elastic-Net} ~\citep{zou_regularization_2005} model, which combines L1 and L2 regularization, providing feature selection and penalizing for multi-collinearity, respectively. Given the higher prevalence of low-uncertainty lesions and a greater interest in explaining high-uncertainty lesions, we implemented an increased penalty for high-uncertainty examples, \textit{i.e.}, sample weighing based on uncertainty in the loss function and the quality metrics. The model selection procedure was performed using grid search, optimizing the weighted \(R^2 \) score obtained on 5-fold cross-validation. The hyperparameters tuned during the model selection control the regularization strength (9 values from $10^{-4}$ to $10^2$ evenly spaced on the log scale), the L1 to L2 regularization strength ratio (9 values evenly spaced between 0 and 1), and the tolerance for the optimization ($10^{-3}$, $10^{-4}$, $10^{-5}$). To obtain robust feature importance values, the grid search procedure was repeated 10 times with different random seed initializations, controlling the data-splitting and \textit{Elastic-Net} convergence. Additionally, to study the relationship between the prediction quality and uncertainty, we used the ordinary least squares regression model. The analysis was repeated 10 times with the random seed change affecting the cross-validation data splitting. 

\textbf{Random forests.} Random forests ~\citep{breiman_random_2001} is an ensemble learning method that builds multiple decision trees during training, using random subsets of data and features, and combines their outputs (via averaging for regression or majority voting for classification) to make predictions. Feature importances in random forests are computed by evaluating how much each feature reduces the impurity (in our regression case, variance) across all trees in the forest. 
The model selection procedure was performed using grid search, optimizing the $ R^{2} $ score obtained on 5-fold cross-validation. The search grid was defined by varying the number of estimators (20, 50, 100), the maximum depth of the trees (none, 5, 10), the minimum number of samples required to split an internal node or to be at a leaf node (2, 5), and the number of considered features for splitting (N, N²). Analogously to the linear model, the grid search was repeated 10 times with different random seed initializations.

\subsubsection{Feature importance}
The coefficients of the linear models were used to evaluate the strength of the relationship between features and uncertainty, given that the features have a similar scale. For each of the datasets, \textit{i.e.}, Train, Test-in, Test-out, the coefficients were obtained during each of 10 runs, corresponding to different random seeds. During each run, the model selection procedure was performed (grid search with 5-fold cross-validation $ R^{2} $ optimization) and the model was fitted on the dataset. Reported feature importances were obtained by averaging across 10 runs, the standard deviation was computed for each selected feature to evaluate the variability.

\subsection{Evaluation of explanations}

\subsubsection{Quality of fit}

To assess the quality of the model fitting during the model selection, the coefficient of determination $ R^{2} $ was used, which represents the proportion of the variance in the dependent variable (\textit{i.e.} LSU) that is explained by the independent variables (\textit{i.e.} lesion or patient features) in a linear regression model:

\begin{equation*} 
R ^ 2 = 1 - \frac{SS_{res}}{SS_{tot}} = 1 -\frac{\sum_{i=0}^{n-1} w_i(y_i- \hat y_i)^{2}}{\sum_{i=0}^{n-1} w_i(y_i- \bar y_i)^{2}},
\end{equation*}

 where $ y_i $, $ \hat y_i $ are true and predicted values, $SS_{res}$ is the residual sum of squares or unexplained variance, $SS_{tot}$ is the total sum of squares or total variance in the observed variable, $\bar{y} = \frac{1}{n} \sum_{i=0}^{n-1} w_i y_i
$, $w_i$ - sample weights. If $ R^{2} $ is unweighed $ w_i = 1, \quad i = 0, \dots, n - 1 $.

 $R^{2} = 1 $ indicates the best possible quality of fit, and $ R^{2} <0 $ indicates that the model is worse than using the mean of the dependent variable y as a prediction.
Additionally, the error magnitude was reported using the mean absolute error (MAE):

\begin{equation*}
    MAE = \frac{1}{n} \sum_{i=1}^{n} w_i |y_i - \hat{y}_i|
\end{equation*}

The quality of fit, used as a proxy for explanation quality, was computed over 10 independent runs—either within the 5-fold cross-validation setting or across dataset splits (Train, Test-in, Test-out) for the best-selected model. We report the mean and standard deviation across runs initialized with different random seeds. We conducted non-parametric statistical tests to assess the significance of the fitting quality differences across models (paired Wilcoxon signed-rank test) and datasets (Mann-Whitney U-test).

\subsubsection{Expert rater feedback}
\label{rater-feedback}

\begin{figure}[!h]
    \centering
    \caption{The clinical questionnaire with guidelines provided to the doctors on the left and questions on the right.}
    \includegraphics[width=\linewidth]{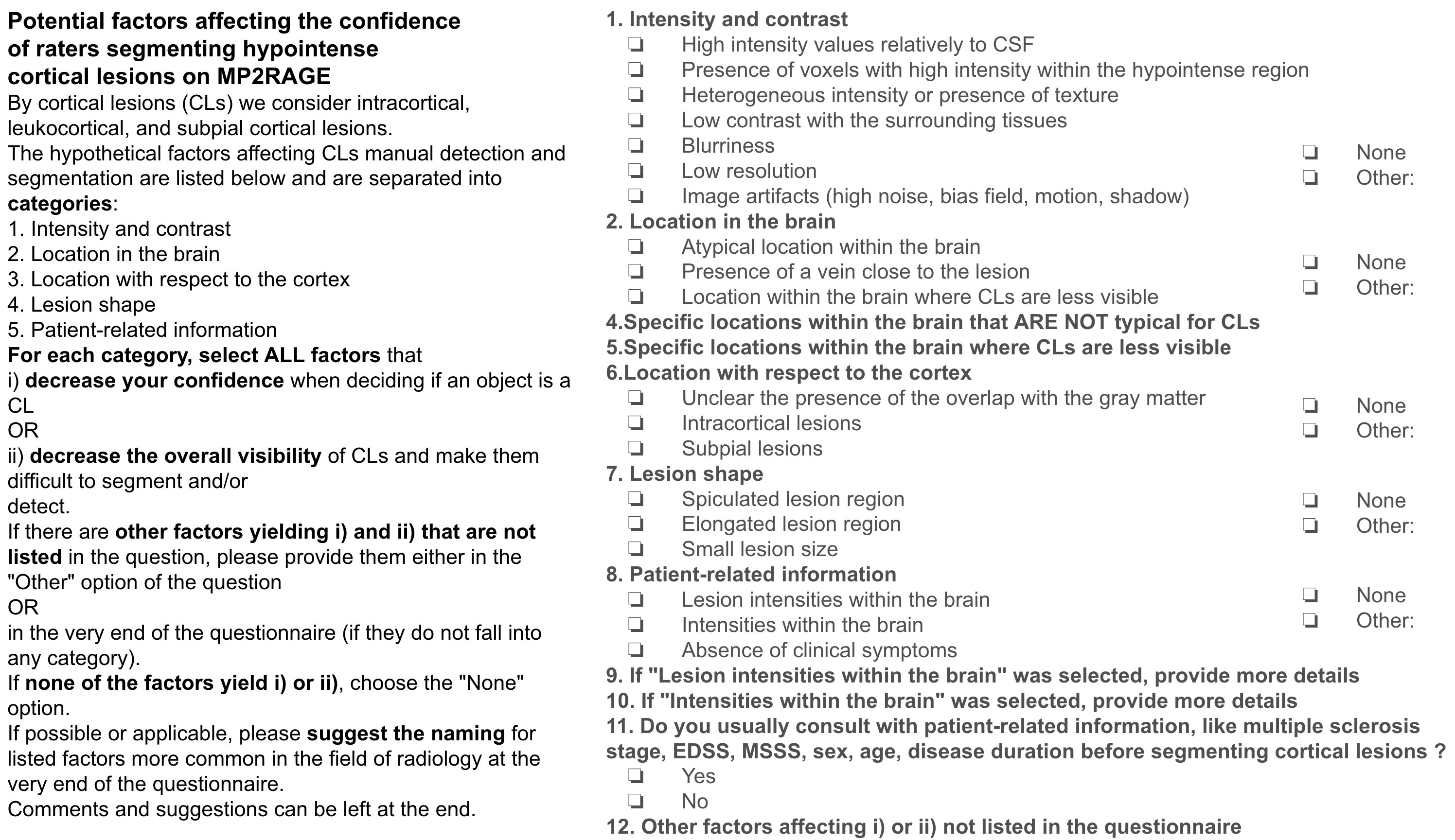}
    \label{fig:questionnaire}
\end{figure}

We used feedback from the experts annotating CLs to study the relationship between the factors underlying model uncertainty and the challenges in manual CL annotation. The questionnaire was used as structured qualitative triangulation to assess whether the factors associated with predictive uncertainty align with clinician-reported sources of doubt during CL annotation, rather than as a formal clinical validation study. We aimed to identify the factors that may hinder medical doctors’ confidence in annotating CLs. We assume that any factor should lie in one of five categories: i) intensity and contrast, ii) location in the brain, iii) location with respect to the cortex, iv) lesion shape, and v) patient-related information. Using the domain knowledge, available guidelines, our prior research, and unstructured interviews with doctors, we proposed several factors for each category, leaving the possibility for the expert raters to suggest other factors. For each of the categories, the respondents needed to select all the factors that decrease either confidence in deciding if an object is a CL or the overall visibility of CLs, making them difficult to segment or detect. We left to the respondents the right to reject any of the factors or propose new ones. All the questions can be found in Figure \ref{fig:questionnaire}. In total, five medical doctors with different levels of expertise underwent the questionnaire: one neuroscientist with 3 years of experience and 4 neurologists with 6, 9, 10, and 20 years of experience in MS.

\section{Results}

\subsection{Deep learning segmentation model performance}

Table \ref{tab:perf} reports CL segmentation and detection performance of the trained nnU-Net model. The segmentation and lesion detection performance are significantly worse on Test-out compared to Train ($\textit{P}< 0.01$) and Test-in ($\textit{P} < 0.05$ for Precision and $\textit{P} < 0.01$ for other metrics). This enables the evaluation of uncertainty explanations under a strong domain shift leading to model failure.

Table \ref{tab:perf} also reports the Spearman's correlation between uncertainty and error per dataset (Unc.-error $\rho$). For Train and Test-in, LSU shows a correlation above 0.5 with lesion-wise prediction quality, whereas on Test-out $\rho$ decreases to 0.35, signaling a weaker uncertainty-error relationship out-of-domain.

The number of predicted CLs, \textit{i.e.}, the number of examples used to fit the explainer model, per dataset is 753 lesions for Train, 284 for Test-in, and 273 for Test-out. 

All P-values from the statistical tests are provided in the Supplementary Materials.

\begin{table}[h!]
\centering
\caption{Mean and standard deviation of the segmentation (DSC, nDSC) and lesion detection quality metrics (LF1, LPPV, LTPR) across Train, Test-in, and Test-out datasets, as well as the number of predicted CL (\# CLs) and correlation between LSU and IoU\textsubscript{adj} (Unc.-error $\rho$).}
\renewcommand{\arraystretch}{1.3}
\small
\begin{tabular}{|p{1.5cm}|*{7}{p{1.3cm}|}}
\hline
\textbf{Dataset} & \textbf{DSC} & \textbf{nDSC} & \textbf{LF1} & \textbf{LPPV} & \textbf{LTPR} & \textbf{\# CLs} & \textbf{Unc.-error} $\mathbf{\rho}$ \\
\hline
Train & 0.908 $\pm$ 0.082 & 0.740 $\pm$ 0.289 & 0.957 $\pm$ 0.078 & 0.991 $\pm$ 0.030 & 0.934 $\pm$ 0.120 & 753 & 0.543 \\
\hline
Test-in & 0.617 $\pm$ 0.332 & 0.494 $\pm$ 0.380 & 0.728 $\pm$ 0.300 & 0.759 $\pm$ 0.333 & 0.778 $\pm$ 0.287 & 284 & 0.545 \\
\hline
Test-out & 0.318 $\pm$ 0.373 & 0.284 $\pm$ 0.372 & 0.399 $\pm$ 0.396 & 0.529 $\pm$ 0.464 & 0.377 $\pm$ 0.406 & 273 & 0.353 \\
\hline
\end{tabular}
\label{tab:perf}
\end{table}

\subsection{Performance in explaining uncertainty}
\label{performance-expl}

The quality of fit is presented in Table \ref{tab:4} for linear models. The contribution of different feature subgroups was investigated, including i) all features; ii) only prediction quality measured by IoU\textsubscript{adj}; iii) only lesion-scale features, \textit{i.e.}, all the patient-scale features were excluded; and iv) all features excluding prediction quality. 

For the linear model, using all features, the \(R^2 \) was highest when fitted on the Train set; it decreased by 20\% on Test-in and by an additional 33\% on Test-out. The MAE increased by 50\% from Train to Test-in and by 17\% from Test-in to Test-out. These drops in the performance across the domains are statistically significant according Mann-Whitney U-test with $\textit{P}<0.001$. The standard deviation of \(R^2 \) on the Test-out set is high due to an outlier appearing under a specific random seed. However, this outlier did not affect the feature importance analysis (see Supplementary materials). 

The quality of fit, measured by \(R^2 \), more than doubled when all features were used compared to using only lesion segmentation quality to predict uncertainty. The change is statistically significant according to the paired Wilcoxon tests ($\textit{P}<0.01$) for all datasets. This suggests that the designed features captured more variability in uncertainty than lesion segmentation quality (IoU\textsubscript{adj}) alone. Fitting the explainer on all features, but the predicted quality, led to up to an 11\% decrease in the \(R^2 \) compared to using all features on Test-in and Test-out ($\textit{P}< 0.01$), but no significant change on Train. This indicates that the proposed features could not completely replace the information about the segmentation error. When patient scale information was omitted, \(R^2 \) improved by 2\% on Train ($\textit{P}=0.006$) and 3\% on Test-out ($\textit{P}=0.431$), but decreased by 2\% on Test-in ($\textit{P}=0.232$).

\begin{table}[h!]
\centering
\caption{The quality of fit of the linear models measured by the coefficient of determination $R^2$ ($\uparrow$) and $MAE$ ($\downarrow$) computed on cross-validation on Train, Test-in, and Test-out sets. The mean and standard deviation are computed across 10 runs with different random seeds. Different feature spaces are explored, including all features, only lesion-scale features, and all features without prediction quality. Best values per column are highlighted in \textbf{bold}.}
\renewcommand{\arraystretch}{1.3}
\small
\begin{tabular}{|p{2cm}|*{6}{p{1.5cm}|}}
\hline
\textbf{Feature set} & \multicolumn{2}{c|}{\textbf{Train}} & \multicolumn{2}{c|}{\textbf{Test-in}} & \multicolumn{2}{c|}{\textbf{Test-out}} \\
\cline{2-7}
& \(\mathbf{ R ^ 2 }\) & \textbf{MAE} & \(\mathbf{ R ^ 2 }\) & \textbf{MAE} & \(\mathbf{ R ^ 2 }\) & \textbf{MAE} \\
\hline
All & 0.571 $\pm$ 0.039 & 0.088 $\pm$ 0.001 & \textbf{0.465 $\pm$ 0.046} & \textbf{0.132 $\pm$ 0.003} & 0.309 $\pm$ 0.202 & 0.155 $\pm$ 0.005 \\
\hline
Only IoU\textsubscript{adj} & 0.237 $\pm$ 0.062 & 0.125 $\pm$ 0.001 & 0.227 $\pm$ 0.067 & 0.168 $\pm$ 0.003 & -0.012 $\pm$ 0.252 & 0.192 $\pm$ 0.005 \\
\hline
Only lesion-scale & \textbf{0.580 $\pm$ 0.039} & \textbf{0.087 $\pm$ 0.001} & 0.458 $\pm$ 0.043 & 0.137 $\pm$ 0.004 & \textbf{0.318 $\pm$ 0.181} & \textbf{0.153 $\pm$ 0.005} \\
\hline
No IoU\textsubscript{adj} & 0.568 $\pm$ 0.037 & 0.089 $\pm$ 0.001 & 0.414 $\pm$ 0.044 & 0.139 $\pm$ 0.005 & 0.278 $\pm$ 0.249 & 0.160 $\pm$ 0.006 \\
\hline
\end{tabular}
\label{tab:4}
\end{table}

The random forest quality of fit is shown in Table \ref{tab:5}. The MAE values of random forest models are comparable to those of the linear models for Train and Test-in. The differences in MAE on Train and Test-in are not statistically significant according to the Wilcoxon rank test ($\textit{P}>0.1$) for all settings, but Train with only lesion-scale features ($\textit{P}=0.006$). The \(R^2 \) is lower than the linear model for each feature set: up to 15\% lower on Train and Test-in ($\textit{P}<0.01$), up to 60\% lower on Test-out ($\textit{P}<0.1$). For all datasets, removing either patient-scale information or the prediction quality led to lower \(R^2 \) values compared to using all features. However, the difference in \(R^2\) is statistically significant only when removing IoU\textsubscript{adj} and fitting on Test-in and when removing patient information and fitting on Test-out ($\textit{P} < 0.01$). Despite slightly lower performance, the standard deviation of the cross-validation scores across different seeds is lower for the random forest model than for linear models.

\begin{table}[h!]
\centering
\caption{The quality of fit of the random forest models measured by the coefficient of determination $R^2$ (↑) and $MAE$ (↓) computed on cross-validation on Train, Test-in, and Test-out sets. The mean and standard deviation are computed across different random seeds. Different feature spaces are explored, including all features, only lesion-scale features, and all features without prediction quality. Best values per column are highlighted in \textbf{bold}.}
\renewcommand{\arraystretch}{1.3}
\small
\begin{tabular}{|p{2cm}|*{6}{p{1.5cm}|}}
\hline
\textbf{Feature set} & \multicolumn{2}{c|}{\textbf{Train}} & \multicolumn{2}{c|}{\textbf{Test-in}} & \multicolumn{2}{c|}{\textbf{Test-out}} \\
\cline{2-7}
& \(\mathbf{ R ^ 2 }\) & \textbf{MAE} & \(\mathbf{ R ^ 2 }\) & \textbf{MAE} & \(\mathbf{ R ^ 2 }\) & \textbf{MAE} \\
\hline
All & \textbf{0.504 $\pm$ 0.031} & \textbf{0.089 $\pm$ 0.003} & \textbf{0.398 $\pm$ 0.036} & \textbf{0.135 $\pm$ 0.005} & \textbf{0.184 $\pm$ 0.084} & \textbf{0.167 $\pm$ 0.006} \\
\hline
Only lesion-scale & 0.497 $\pm$ 0.025 & 0.090 $\pm$ 0.002 & 0.391 $\pm$ 0.040 & 0.136 $\pm$ 0.005 & 0.121 $\pm$ 0.084 & 0.172 $\pm$ 0.008 \\
\hline
No IoU\textsubscript{adj} & 0.495 $\pm$ 0.025 & 0.090 $\pm$ 0.003 & 0.356 $\pm$ 0.035 & 0.141 $\pm$ 0.003 & 0.181 $\pm$ 0.086 & 0.168 $\pm$ 0.007 \\
\hline
\end{tabular}
\label{tab:5}
\end{table}

\begin{figure}[!h]
\caption{The quality of fit of the linear models measured by the coefficient of determination 
\(R^2 \) ($\uparrow$, red) and MAE ($\downarrow$, blue) computed on Train, Test-in, and Test-out sets. Rows correspond to different pairs of sets used for fitting a linear model and different feature sets; columns correspond to the sets used for the evaluation of the linear model. The mean and standard deviation are computed across different random seeds.}
    \centering
    \includegraphics[width=1\linewidth]{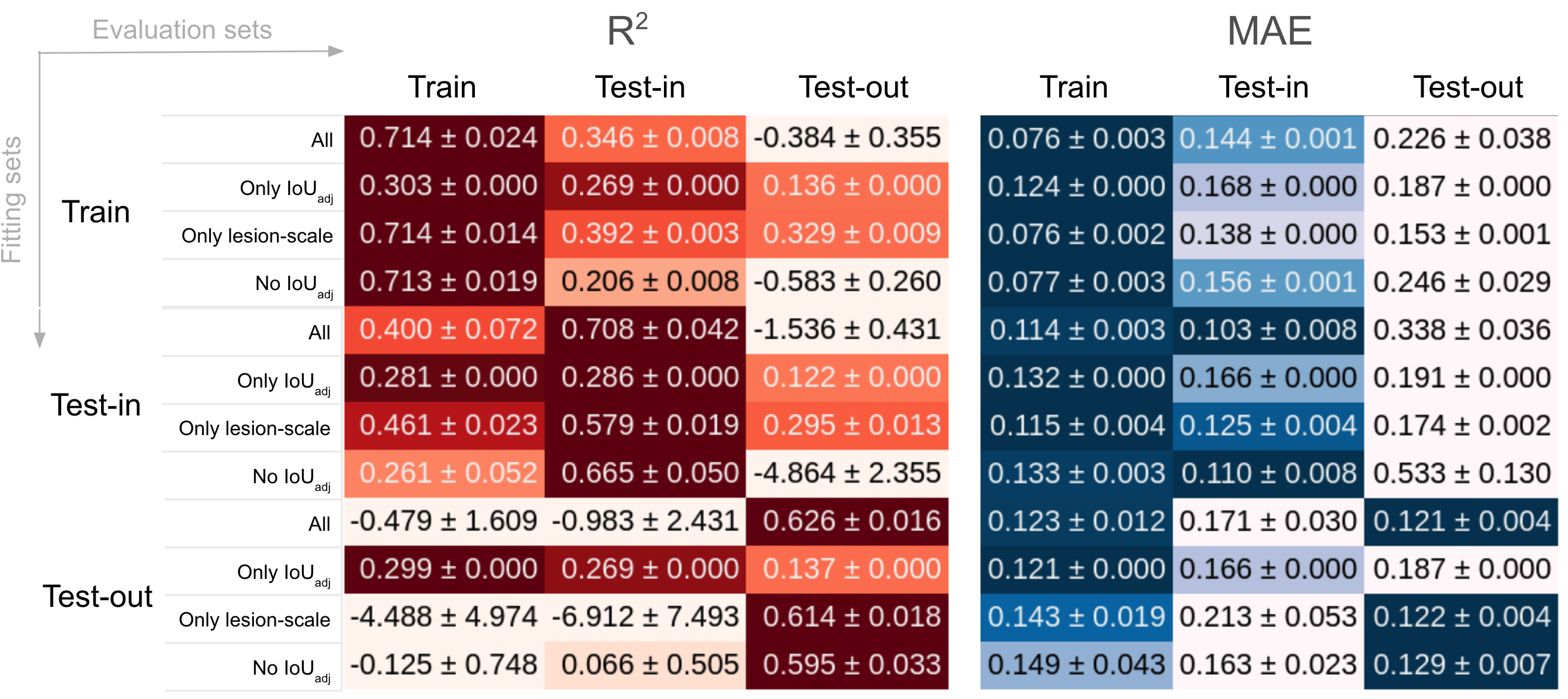}
    \label{fig:heatplot}
\end{figure}

Given these results, we chose the ElasticNet model with all features as the main explanatory model. We analyzed this model prediction quality on sets different from the fitting sets and evaluated the quality of explanations under the domain shift. The results are shown in Figure \ref{fig:heatplot}. A drop in performance caused by the domain shift was observed for all fitting settings, except for one. When IoU\textsubscript{adj} was used to predict uncertainty using Test-out for fitting, the \(R^2 \) was higher on Train and Test-in than on Test-out, indicating a weaker relationship between uncertainty and error on the out-of-domain set, compared to in-domain sets. Overall, the drop in performance tended to be worsened by the out-of-domain setting, and the differences in the performance are statistically significant with $\textit{P}<0.01$ according to the Mann-Whitney U-test. Thus, explanations on the Train and Test-in sets should be more similar to each other than to Test-out.

The exact $\textit{P}$-values for all statistical tests are provided in the Supplementary material.

\subsection{Feature importance analysis}
\label{results-importance}

\begin{figure}[h!]
    \centering
    \caption{Barplot with linear explainer coefficients treated as feature importance. Barplots are built for Train, Test-in, and Test-out sets. Positive values - direct relationship with uncertainty, and negative values - direct relationship with certainty. Twenty features with the highest absolute coefficients are built. MAD -mean absolute deviation.}
    \includegraphics[width=0.92\linewidth]{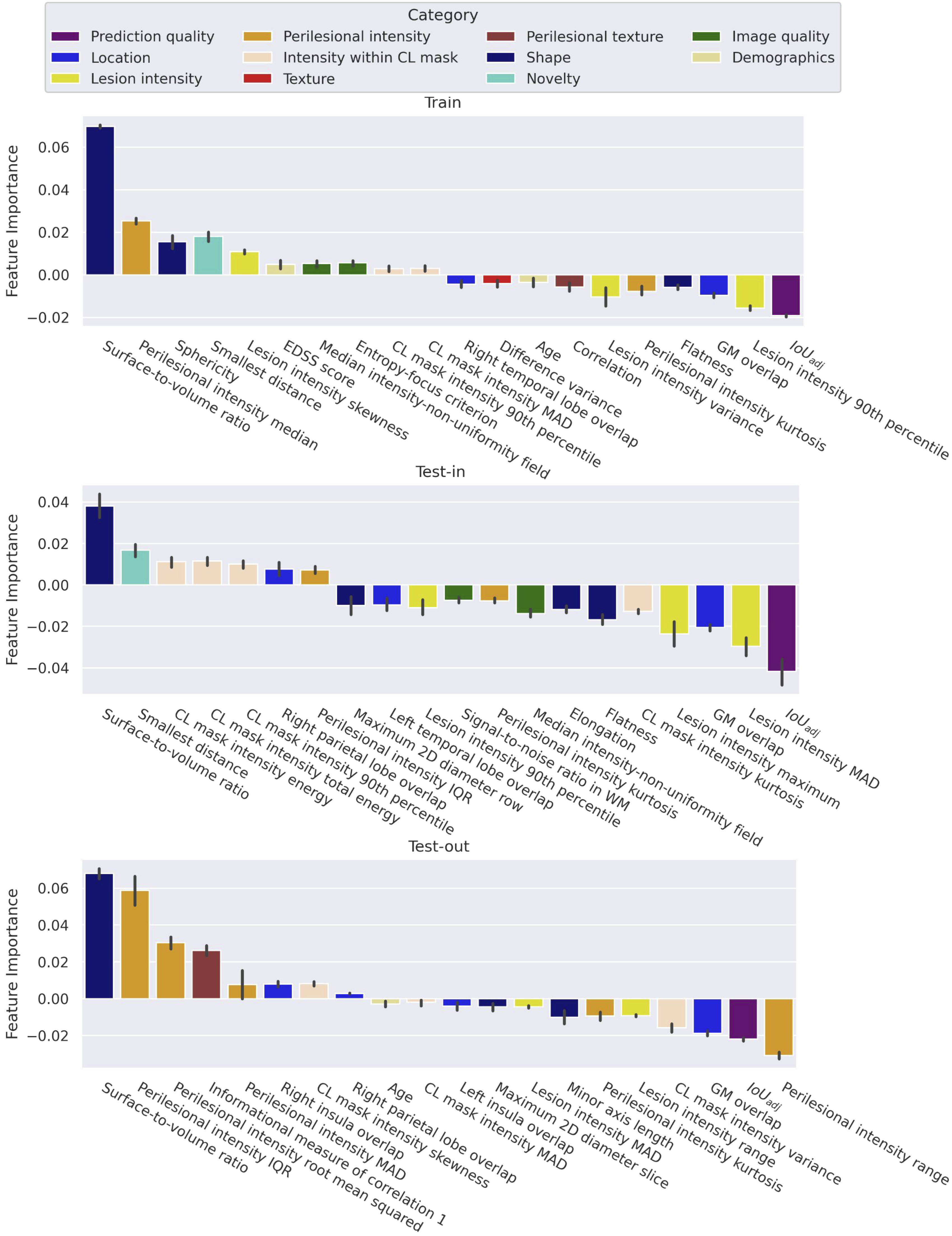}
    \label{fig:elasticnet_importances}
\end{figure}

Figure \ref{fig:elasticnet_importances} presents coefficients of the \textit{Elastic-Net} model fitted to the entire feature space, ranked by magnitude; 20 coefficients explaining the biggest part of the variance are displayed. Since all the features were standardized before model fitting, the absolute values are comparable and indicate feature importance. The features with the highest absolute importance explain the most variability in uncertainty. Positive values of coefficients indicate features that correlate with increased uncertainty (when the feature value rises, uncertainty typically increases), while negative coefficients represent features associated with decreased uncertainty.  

There were similarities in the selected features between different fitting sets (Train, Test-in, and Test-out). The surface-to-volume ratio explained the variability in uncertainty across all fitting sets and corresponded to higher uncertainty values. The surface-to-volume ratio describes both the shape and size of the object. High values of this feature correspond to irregular or elongated shapes and smaller sizes of the objects. Other features describing shape and size emerged as relevant predictors of lesion uncertainty for the different datasets. Especially on the test sets, more spherical lesions (high flatness and elongation \textit{PyRadiomics} features and low surface-to-volume ratio) of bigger sizes (high maximum 2D diameter and minor axis length, low surface-to-volume ratio) were related to higher certainty. Prediction quality measured by IoU\textsubscript{adj} was another high-importance feature shared by all the fitting sets, with a higher IoU\textsubscript{adj} value corresponding to lower lesion uncertainty. GM overlap appears in the analysis for all sets. Higher values of GM overlap coincided with lower uncertainty. Lesion novelty feature (smallest distance) appeared for both Train and Test-in sets and had a mediocre importance, indicating that lesions dissimilar to the Train set lesions have higher uncertainty. Different intensity features were selected during the feature importance analysis. They are related not only to the lesion region itself, but also to the lesion surrounding, and the whole predicted lesion region. The particular features selected vary between different sets. Generally, the presence of (outlying) high intensity within a lesion indicated lower uncertainty (high lesion intensity 90\% percentile, variance, mean absolute deviation, maximum, range). Perilesional intensity kurtosis was selected for all fitting sets, however, it had a low relative importance. Higher perilesional intensity kurtosis was associated with lower uncertainty values. Some feature groups were not selected or have relatively low importance, including texture, location in the brain, patient characteristics, and image quality (except for the Test-in set). While Test-in and Test-out had some similarities in the selected features, there were some visible differences. First, a weaker relationship between uncertainty and error in Test-out, judging by the lower relative importance of IoU\textsubscript{adj}. Additionally, for Test-out, the accumulated importance of the perilesional intensity features was higher than for shape features or IoU\textsubscript{adj}. The feature importance plots for the random forest model are provided in the Supplementary Materials.

\subsection{Selected features analysis}
\label{select-feature-analysis}

\begin{figure}[!h]
    \centering
    \caption{Bivariate histograms with joint distributions of features and lesion uncertainty for surface-to-volume ratio, IoU\textsubscript{adj}, GM overlap, and (lesion novelty) smallest distance features computed on Train, Test-in, and Test-out data. Pearson correlation $r$ is computed on the standardized features.}
    \includegraphics[width=.75\linewidth]{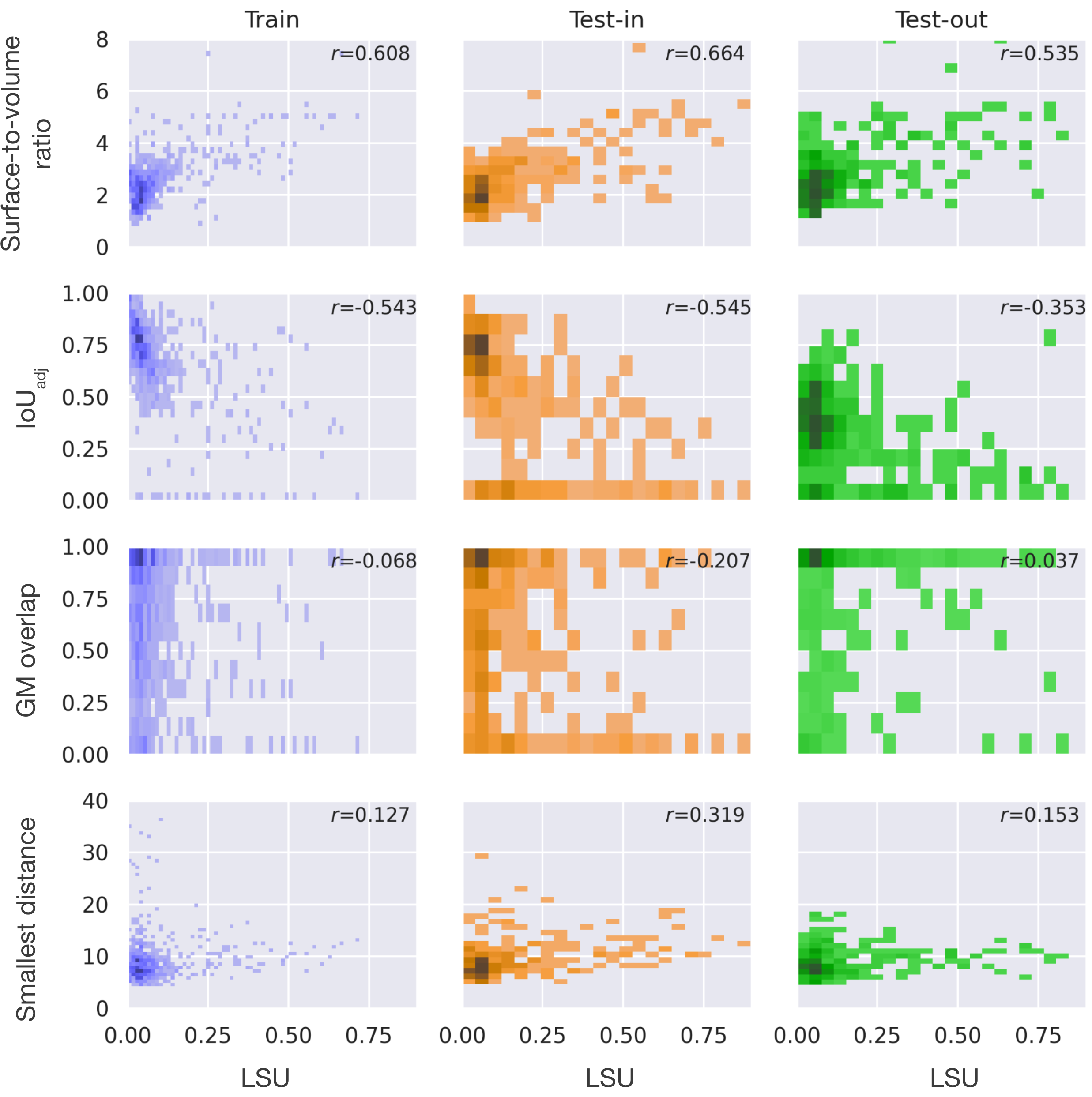}
    \label{fig:individual-features}
\end{figure}

The joint distribution of uncertainty and high-importance features in different datasets is shown in Figure \ref{fig:individual-features}. The median Pearson correlation between uncertainty and surface-to-volume ratio is 0.61, and IoU\textsubscript{adj} is 0.54. Compared to the in-domain sets, the correlation with IoU\textsubscript{adj} is 35\% lower on the out-of-domain Test-out. Thus, the relationship between uncertainty and lesion prediction quality, measured by IoU\textsubscript{adj}, is weaker on the out-of-domain set. GM overlap had a weak correlation with lesion uncertainty (r=-0.07 for Train, -0.21 for Test-in, and -0.04 for Test-out). The shape of the distribution indicates that the relationship is not linear. Lesion novelty measured by the \textit{Smallest distance} feature had a weak positive correlation ranging from 0.13 to 0.32 Pearson correlation coefficient.

\begin{figure}[!h]
    \centering
    \caption{Lesion examples sampled based on the feature distribution. The shape, location, and intensity features were selected based on the feature importance analysis. CLs were obtained based on 0, 25, 50, 75, and 100 percentiles (columns from left to right) of each feature distribution on a particular set (rows - Train, Test-in, or Test-out). All the features have a positive relationship with lesion-wise uncertainty, except for GM overlap and lesion intensity mean absolute deviation, which are negatively related to uncertainty. Only lesions with a non-zero overlap with the ground truth are displayed.}
    \includegraphics[width=0.8\linewidth]{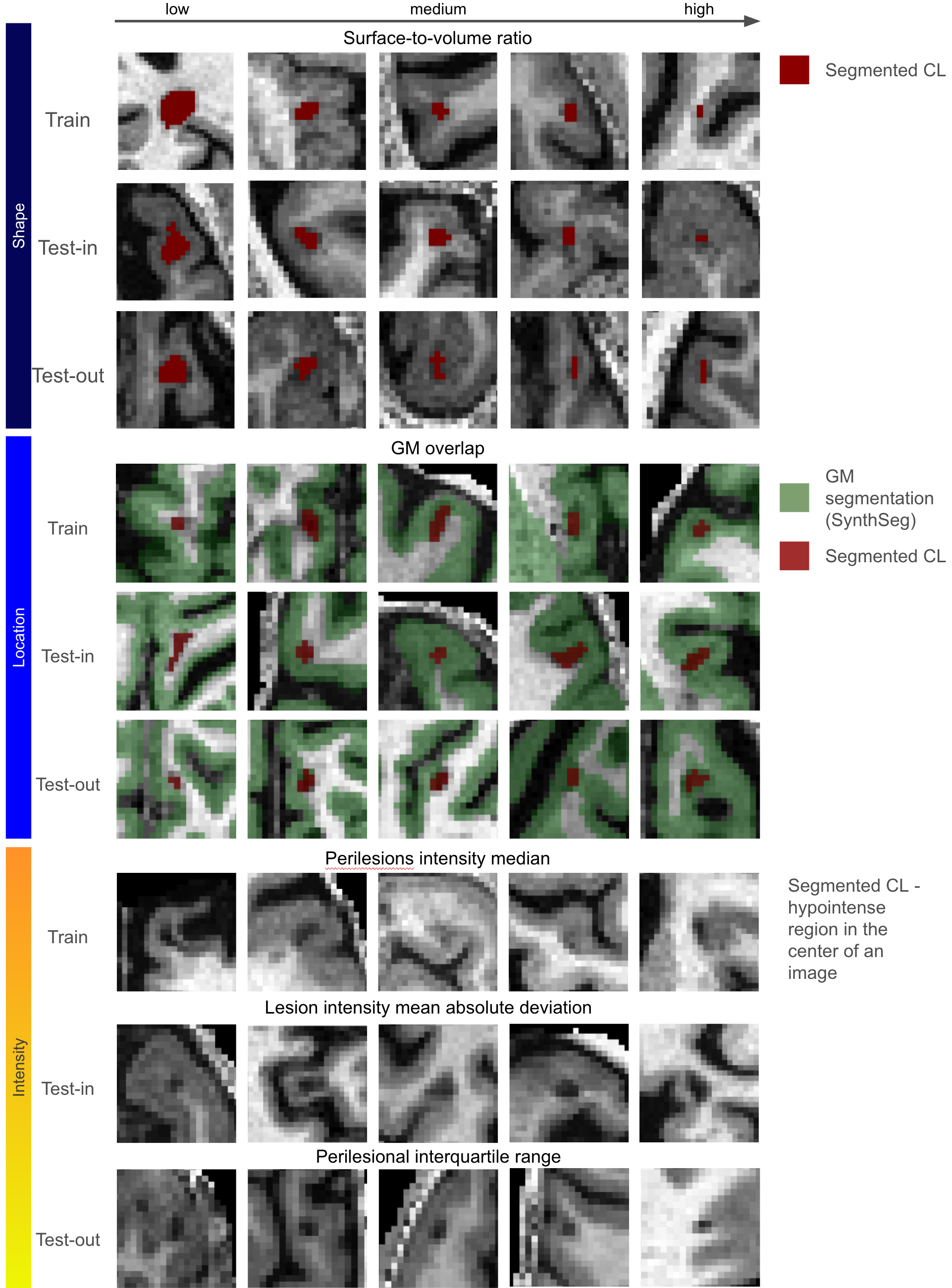}
    \label{fig:examples}
\end{figure}

To better understand the information contained in features, we used lesion visualization. Figure \ref{fig:examples} shows lesion examples obtained based on the feature distributions for shape, location, and intensity high-importance features. The feature with the highest importance, \textit{i.e.}, surface-to-volume ratio, clearly communicates the lesion size; lesion elongation could also be captured by this feature. The GM overlap feature, important for all sets, is reflective of the juxtacortical (a WML, adjacent to the cortex, but contained in WM), leukocortical (a CL, overlapping with GM and WM), and intracortical lesions (contained in GM). This said, the segmentation of GM is subject to errors and partial volume effect, potentially impairing the reliability of the derivative feature. We find that all the highest-importance intensity features carried some information about the location with respect to GM. The perilesional intensity median indicated a greater border with WM while being associated with higher uncertainty. Thus, this feature could be a proxy for the juxtacortical versus CL distinction. Similarly, the mean absolute deviation of the lesion intensity and perilesional intensity IQR possibly served this distinction.

\subsection{Expert rater questionnaire}

\begin{figure}[h!]
    \centering
    \caption{Barplot summarizing the expert rater feedback completed by five MP2RAGE CL annotators. The different colors correspond to different categories of features from the questionnaire.}
    \includegraphics[width=\linewidth]{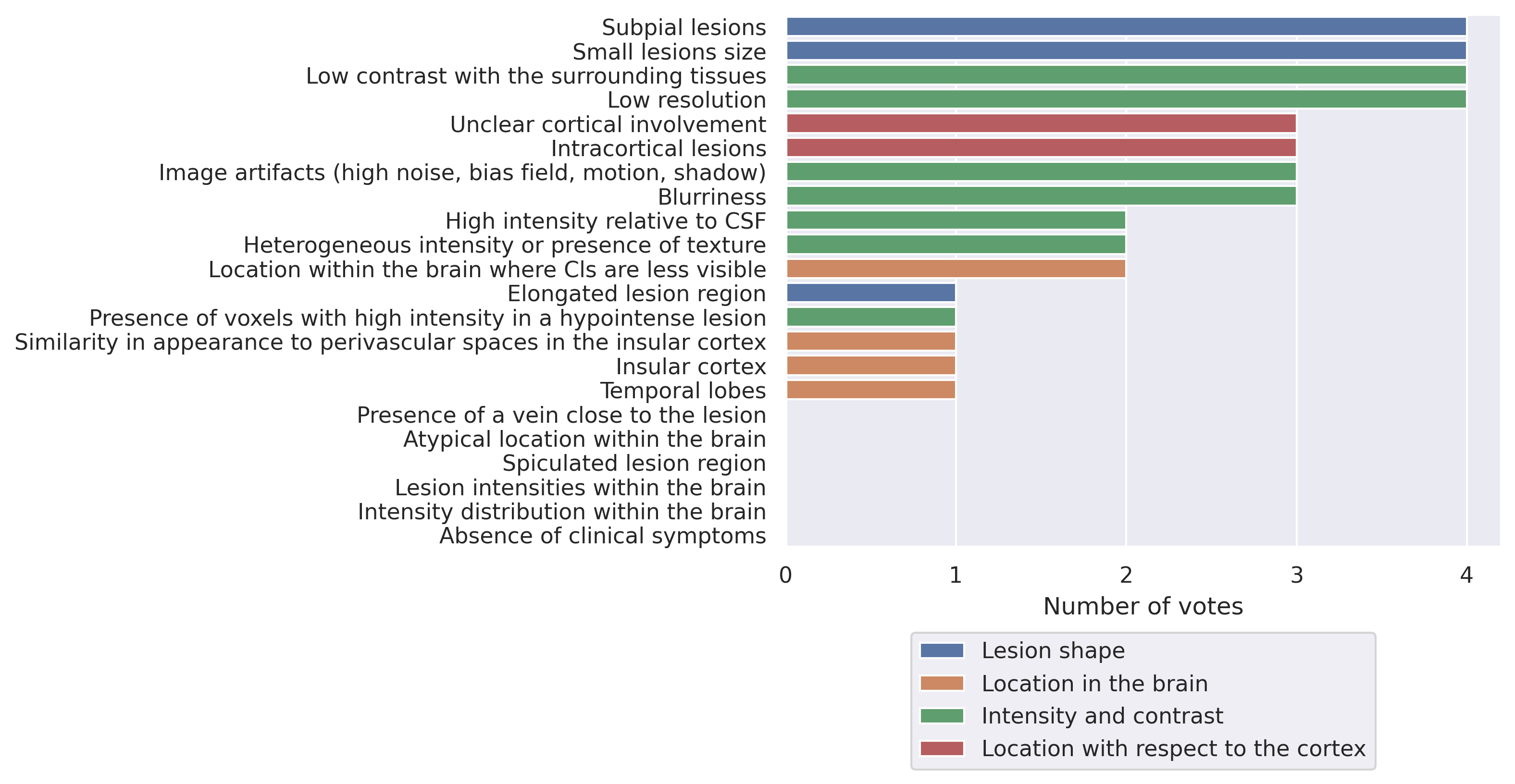}
    \label{fig:quest-res}
\end{figure}

Figure \ref{fig:quest-res} summarizes the results of the expert rater questionnaire, providing qualitative support for the clinical plausibility of the identified uncertainty factors by highlighting key contributors to low visibility and reduced manual segmentation quality of CLs on MP2RAGE. The proposed concepts were additioned by the respondents with respect to the "Location in the brain" category. The proposed concepts include i) similarity in appearance to perivascular spaces in the insular cortex and ii) temporal lobes.

The responses indicated that subpial lesions, small lesions, lesions with low contrast with the surrounding tissue, and scans with low resolution are the most challenging to annotate (4 out of 5 raters). Lesion location relative to the cortex was also selected by the majority of participants, suggesting that both lesions with unclear cortical involvement and intracortical lesions pose significant challenges for annotators. Image quality-related factors, such as artifacts and blurriness, were also selected by the majority of respondents as additional obstacles. Furthermore, the presence of high-intensity values or heterogeneous intensity patterns within a lesion was noted. Two experts specifically highlighted the insular cortex and temporal lobes as brain regions where CLs are more difficult to annotate. None of the respondents selected patient-related factors as influencing annotation difficulty, and all reported that patient-specific information was not accessed before annotation.

\section{Discussion}

The proposed analysis framework, which regresses uncertainty from human-interpretable features, explores the information encoded by instance-wise deep ensemble uncertainty for multiple sclerosis cortical lesion segmentation on MRI.

\subsection{Information carried by instance-wise uncertainty}

Our findings demonstrated that deep ensemble uncertainty is not solely explained by the quality of the prediction (IoU\textsubscript{adj}) or distance from the training set (\textit{i.e.}, lesion novelty). Instead, various features characterizing the size, shape (surface-to-volume ratio), and location of the lesion with respect to the cortex (GM overlap and various intensity features) were more important in explaining the variability of uncertainty. These factors align with the main factors reported by expert raters as reducing confidence during CL annotation, providing qualitative triangulation for the clinical plausibility of the findings: small and elongated (subpial) lesions and the distinction between juxtacortical and CL (\textit{i.e.}, cortical involvement). 

This alignment between model uncertainty and annotator uncertainty suggests that deep ensemble uncertainty also encodes aleatoric noise. Deep ensemble uncertainty is commonly associated with epistemic uncertainty and, thus, more strongly correlated with model error than other UQ methods ~\citep{lakshminarayanan_simple_2017, ovadia_can_2019}. However, this assumption stems largely from low-noise tasks. In contrast, CL segmentation is inherently noisy. Moreover, hidden data biases, such as underrepresentation of certain lesion subtypes (\textit{e.g.}, subpial or small intracortical lesions) ~\citep{cagol_diagnostic_2024, beck_cortical_2022}, may also contribute to increased uncertainty. This interplay of factors raises questions about the separability of the sources of aleatoric and epistemic uncertainty ~\citep{valdenegro-toro_deeper_2022}.

As discussed in Section \ref{select-feature-analysis}, non-linear dependencies are present between uncertainty and selected features with high importance. Partially, an unexplained uncertainty should be attributed to the non-linear relationship. However, modelling these dependencies requires more data, since random forest explainers did not outperform linear models in goodness of fit, likely due to overfitting in small datasets ~\citep{hastie_elements_2009}. While linear models are pragmatic for limited data, their results require supplementary analysis (\textit{e.g.}, partial dependence plots) to characterize non-linear dependencies. 

\subsection{Explainability of uncertainty under domain shifts}

We evaluated uncertainty and the proposed framework performance under domain shift — a clinically realistic scenario where prediction quality varies across settings. As shown in Section \ref{performance-expl}, the explanatory power of uncertainty degraded progressively from Train to Test-in to Test-out, indicating reduced informativeness with growing domain shift. We hypothesize that this degradation is caused by changes in relevant features and a deterioration in uncertainty quality. Prior work already showed that UQ reliability declines under distributional shifts ~\citep{ovadia_can_2019, malinin_shifts_2022a, malinin_shifts_2022b}. 

Our results further suggest that uncertainty estimates under strong domain shift may lose informativeness about both prediction error and aleatoric noise, given higher unexplained uncertainty on Test-out (Tables \ref{tab:4} and \ref{tab:5}). We also observed performance degradation when an explainer was fitted to one set and applied to another. The drop in performance quantified the transferability of explanations between Train and Test-in sets (\textit{i.e.}, in-domain sets), which is better than the in-domain versus Test-out. These findings suggest that uncertainty explainability is sensitive to domain shift, both in terms of uncertainty quality and explainer reliability.

\subsection{Predictive uncertainty and annotator confidence}

We found a strong correspondence between the features associated with high predictive uncertainty (Section \ref{results-importance}) and the factors underlying low annotator confidence (Section \ref{rater-feedback}). An increased surface-to-volume ratio characterizes small lesions and subpial lesions (elongated, extending along the cortex). As shown in Sections \ref{results-importance}, the surface-to-volume ratio feature was associated with higher lesion uncertainty. GM overlap (\textit{i.e.}, cortical involvement) and intensity-based features were relevant for both annotator confidence and model prediction certainty. 

However, a direct one-to-one mapping between clinically identified factors and engineered lesion features is not feasible. The importance of lesion location, as emphasized by experienced raters, also extends to predictive uncertainty. The non-linear relationship between lesion uncertainty and GM overlap could be explained by the fact that the model is both uncertain on lesions with unclear cortex involvement (GM overlap  0) and on the intracortical lesions (GM overlap = 1). Translating engineering characteristics into understandable concepts is not straightforward ~\citep{gerdes_role_2024} and beyond the scope of this work. 

\subsection{Future implications of uncertainty explanations}

Understanding and explaining uncertainty is important across several applications. Technically, identifying ambiguous or difficult cases can guide efforts to improve model reliability and calibration~\citep{lambert_trustworthy_2024}. For clinical users and other stakeholders, such explanations may enhance transparency and support more informed interaction with automated decision-support tools~\citep{lekadir_future-ai_2025}. By relating uncertainty to clinically meaningful features and presenting this information through visualizations, the framework supports interpretation of model behavior by medical professionals. Structured expert feedback further provides qualitative triangulation about which model-associated factors are also perceived by clinicians as sources of doubt.

Our findings indicate that a portion of predictive uncertainty cannot be explained by the examined features and may arise from UQ limitations and approximation errors. Such unexplained uncertainty may be uninformative and could mislead annotators or reduce confidence. Better identification of these UQ failures may improve trust calibration between perceived and actual reliability~\citep{zhang_effect_2020}. 

Although this study is centered on lesion segmentation in brain MRI, the analysis may also be informative for related uncertainty-aware imaging settings. Longitudinal data could help assess the temporal stability of lesion-scale uncertainty and whether the same uncertainty-associated factors remain relevant over time. The current analysis is correlational; future work may also explore causal relationships behind uncertainty estimates. Relating uncertainty-associated lesion factors to downstream clinical endpoints such as disability would require dedicated harmonized clinical studies, especially because imaging biomarkers in MS often show only modest associations with such outcomes.

\section{Conclusion}

This study addresses a core challenge in trustworthy AI for medical imaging: how to make uncertainty estimates understandable and clinically meaningful. We presented an analysis framework that links lesion-level uncertainty to patient- and lesion-specific characteristics, moving beyond traditional error-based evaluation.

Our results show that uncertainty carries clinically relevant information and is sensitive to domain shift. We also identify cases of unexplained uncertainty that could mislead users. Overall, the framework offers a clinically grounded analysis template for related uncertainty-aware imaging studies.

\section*{Code and data availability}

Code for the model implementation and uncertainty analysis is publicly available at \url{https://github.com/NataliiaMolch/interpret-lesion-unc}. Architecture details are described in the repository. A Zenodo record for the trained model weights is available at \url{https://doi.org/10.5281/zenodo.19485057}. The underlying MRI data and manual annotations are not publicly available because they originate from clinical cohorts and are subject to ethical and privacy restrictions.

\section*{Acknowledgment}

This work was supported by the Hasler Foundation Responsible AI program (MSxplain) and the Research Commission of the Faculty of Biology and Medicine (CRFBM) of UNIL. This research was partially supported by the Intramural Program of NINDS, NIH, USA. We acknowledge access to the facilities and expertise of the CIBM Center for Biomedical Imaging, a Swiss research center of excellence founded and supported by Lausanne University Hospital (CHUV), University of Lausanne (UNIL), École polytechnique fédérale de Lausanne (EPFL), University of Geneva (UNIGE), and Geneva University Hospitals (HUG). The authors wish to thank Dr. Mara Graziani for valuable consultations and discussions at the initial stage of this research.

\appendix
\section{Metrics for model performance evaluation}
\label{app1}

Let TP, FP, FN be the number of true positives, false positives, and false negatives. The quality metrics are defined as follows:

\[
F_1 = \frac{2TP}{2TP + FP + FN},
\]

\[
\text{Precision} = \frac{TP}{TP + FP},
\quad
\text{Recall} = \frac{TP}{TP + FN}.
\]

In the case of segmentation, the F1 score is called the Dice similarity score; in the case of lesion detection, the F1, precision, and recall are called LF1, LPPV, and LTPR, respectively.

Lesion detection metrics rely on TP, FP, and FN lesion definitions. For the lesion detection measures, the lesion regions were defined using the connected component analysis with 26 connectivity. True positive lesions were defined as lesions with a non-zero overlap with the ground truth. False positive lesions are predicted connected components without an overlap with the ground truth. False negative lesions are ground truth lesions without an overlap with the predicted lesion mask.

\vspace{0.5em}

\textbf{Normalized Dice similarity score proposed by Raina et al.\ (2023):}

\[
nDSC = \frac{2TP}{2TP + \kappa \cdot FP + FN}, 
\quad \kappa = h(r^{-1} - 1)
\]

where $h$ is the ratio between the positive and the negative classes, 
$r \in (0, 1)$ is the reference value set to the mean fraction of the positive class. 

In our case, it is the lesion load: \( r = 2 \cdot 10^{-5}. \)

\section{Adjusted intersection over the union definition}
\label{app2}

Let $x$ be an input image and let $\hat{S}_x = \{ \hat{y}_z(x, w) \mid z \in x \}$ be the pixel-wise predicted segmentation, where $\hat{y}_z(x, w)$ is the predicted class at pixel $z$ given model parameters $w$. Define:

\begin{itemize}
  \item $\hat{K}_x$ — the set of connected components (segments) in the predicted segmentation $\hat{S}_x$
  \item $K_x$ — the set of connected components in the ground truth segmentation $S_x$
\end{itemize}

For a predicted segment $k \in \hat{K}_x$, define:
\begin{itemize}
  \item $K' = \bigcup \{ k' \in K_x \mid k' \cap k \neq \emptyset \text{ and } \text{class}(k') = \text{class}(k) \}$ — union of all ground truth components that overlap with $k$ and have the same class
  \item $Q = \{ q \in \hat{K}_x \mid q \cap K' \neq \emptyset \}$ — set of all predicted segments that overlap with $K'$
\end{itemize}

Then, the \textbf{adjusted intersection over union} is defined as:

\[
\mathrm{IoU_{adj}}(k) = \frac{|k \cap K'|}{|k \cup (K' \setminus Q)|}
\]

This formulation corrects the standard intersection over union by not penalizing predicted segments for missing parts of the ground truth that are already covered by other predicted segments of the same class.

\section*{Ethics statement}

Studies involving human data were approved by the local ethics committees; informed consent was obtained from all participants before study entry.

\section*{Declaration of generative AI and AI-assisted technologies in the writing process}

During the preparation of this work, the authors used Grammarly and ChatGPT to correct grammatical and stylistic errors. After using this tool/service, the authors reviewed and edited the content as needed and take full responsibility for the content of the publication.

\bibliographystyle{elsarticle-num-names} 
\bibliography{references-2}

\end{document}